\PassOptionsToPackage{sort&compress}{natbib}

\documentclass[preprint, 10pt]{elsarticle}

\usepackage[english]{babel}
\usepackage[latin1]{inputenc}
\usepackage[T1]{fontenc}

\usepackage[usenames]{color}
\usepackage{amsfonts}
\usepackage{textcomp}
\usepackage{lineno}
\usepackage{hyperref}
\usepackage{adjustbox}

\usepackage{./lib/mathident}

\modulolinenumbers[5]
\journal{Knowledge-Based Systems}

\makeatletter

    \def\mathlistingfont{\mathsf}
    \def\listingfont{\sffamily}

    \def\@code{%
        \begingroup%
            \ifmmode%
                \vbox to 0pt {\hbox to 0pt{\ensuremath{\mathsf{}}}}%
                \bgroup\catcode`\ =\active%
            \else%
                \bgroup%
            \fi
            \@do@code%
    }
    \def\@do@code#1{%
            \ifmmode%
                \@code@bquote{\mathlistingfont{#1}}\@code@equote%
            \else
                \@code@bquote{\listingfont{#1}}\@code@equote%
            \fi
            \egroup%
        \endgroup%
    }

    \def\code{\gdef\@code@bquote{}\gdef\@code@equote{}\@code}
    \def\qcode{\gdef\@code@bquote{\mbox{\normalfont``}}\gdef\@code@equote{\mbox{\normalfont''}}\@code}

    \def\cites#1{\citeauthor{#1}'s \cite{#1}}

    \def\then{\mathrel{\Rightarrow}}

    \def\real{\ensuremath{\mathbb{R}}}

    \def\fset{\mathop{\mathbb{F}}}
    \def\seq{\mathop{\mathrm{seq}}}

    \def\pfun{\mathrel{\not\rightarrow}}
    \def\tfun{\mathrel{\rightarrow}}

    \def\upto{\mathrel{..}}

    \let\save@times=\times
    \def\times{\mathrel{\save@times}}
    \let\save@mod=\mod
    \def\mod{\mathop{\save@mod}}
    \let\save@sum=\sum
    \def\sum{\mathop{\save@sum}}

    \def\customrule{\@ifstar\@customrule@star\@customrule@normal}
    \def\@customrule@star{\vspace{1em}\noindent\hrule\mbox{}\vspace{1em}}
    \def\@customrule@normal{\mbox{}\vspace{1em}\noindent\hrule\mbox{}\vspace{.1em}\noindent\hrule\mbox{}\vspace{1em}}

    \newcounter{inlineemumi}
    \def\inlinenum{%
        \bgroup%
            \setcounter{inlineemumi}{0}%
            \def\@item[##1]{\textit{##1}}%
            \def\item{\stepcounter{inlineemumi}\alph{inlineemumi})~\@ifnextchar[{\@item}{\relax}}%
    }

    \def\endinlinenum{%
        \egroup%
    }

    \def\xitem[#1]{\textit{#1}}

    \newcounter{@preliminary@counter}

    \newcounter{@example@counter}

    \newcounter{@notation@counter}

    \newcounter{@definition@counter}
    
    \def\enddefinition{\relax}

    \def\description{%
        \bgroup%
            \smallbreak%
            \def\@item[##1]{\textit{##1}}%
            \def\item{\par\noindent\hangindent=\parindent\hangafter=1\noindent\@ifnextchar[{\@item}{\relax}}%
    }

    \def\enddescription{%
            \endtrivlist%
            \smallbreak%
        \egroup%
    }

    \def\itemize{%
        \bgroup%
            \smallbreak%
            \def\item{\par\noindent\mbox{--}~}%
    }

    \def\enditemize{%
            \smallbreak%
        \egroup%
    }

    \def\bio#1#2{%
        \bgroup
            \bigbreak%
            \noindent%
            \minipage[b]{0.25\hsize}
                \includegraphics[width=\hsize]{./figs/#1.eps}%
            \endminipage%
            \hfill%
            \minipage[b]{0.70\hsize}%
                {\bfseries #2}%
    }

    \def\endbio{%
            \endminipage%
            \bigbreak%
        \egroup
    }

\makeatletter

\input{./Dirty-Tricks}

\begin{document}


\title{On Extracting Data from Tables \\ that are Encoded using HTML}

\author{Juan C.~Roldán\corref{corresponding}}
\ead{jcroldan@us.es}

\author{Patricia Jiménez}
\ead{patriciajimenez@us.es}

\author{Rafael Corchuelo}
\ead{corchu@us.es}

\cortext[corresponding]{Corresponding author.}
\address{University of Seville, ETSI Informática \\ Avda.~Reina Mercedes s/n, Sevilla E-41012, Spain}


\begin{abstract}
Tables are a common means to display data in human-friendly formats.  Many authors have worked on proposals to extract those data back since this has many interesting applications. In this article, we summarise and compare many of the proposals to extract data from tables that are encoded using HTML and have been published between $2000$ and $2018$. We first present a vocabulary that homogenises the terminology used in this field; next, we use it to summarise the proposals; finally, we compare them side by side. Our analysis highlights several challenges to which no proposal provides a conclusive solution and a few more that have not been addressed sufficiently; simply put, no proposal provides a complete solution to the problem, which seems to suggest that this research field shall keep active in the near future.  We have also realised that there is no consensus regarding the datasets and the methods used to evaluate the proposals, which hampers comparing the experimental results.
\end{abstract}


\maketitle


\section{Introduction}
\label{sec:introduction}

Tables are a common means of displaying data in web documents because people can easily spot and interpret them~\cite{journals/pvldb/CafarellaHWWZ08, journals/pvldb/CafarellaHLMYWW18}.  The estimations are as high as hundreds of millions; for instance, \citet{conf/www/LehmbergRMB16} and \citet{conf/kesw/GalkinMA15} found  $233$ and $12$ billion tables in different editions of the Common Web Crawl, respectively, and \citet{conf/wsdm/CrestanP11} found $8.2$ billion tables in their own crawl. \citet{journals/pvldb/CafarellaHLMYWW18} also highlighted the explosion of consumer demand for data that comes from tables thanks to the increasing popularity of voice assistants and infobox-like search results.

In this context, data extraction consists in transforming tables into structured formats that focus on their data and abstract away from how they are displayed. Data extraction has many applications to text mining~\cite{conf/das/WangH02, conf/ieaaie/FumarolaWBMH11, conf/wecwis/WuCY02}, data (meta-)search~\cite{conf/wecwis/WuCY02, conf/das/WangH02, conf/www/GatterbauerBHKP07, journals/pvldb/CafarellaHWWZ08, journals/pvldb/VenetisHMPSWMW11, journals/pvldb/PimplikarS12, conf/sigmod/ChuHCG15, conf/bdc/EberiusBHTAL15, conf/nldb/MilosevicGHN16}, query expansion~\cite{conf/wsdm/CrestanP11}, document summarisation~\cite{conf/icdcs/LoWY00, conf/das/WangH02}, question answering~\cite{conf/wecwis/WuCY02, journals/vldb/ElmeleegyMH11, conf/er/BraunschweigTL15, conf/nldb/MilosevicGHN16, conf/aaai/NishidaSHM17}, knowledge discovery~\cite{journals/eaai/KimL05, conf/www/GatterbauerBHKP07, conf/sigmod/ChuHCG15, conf/nldb/MilosevicGHN16, journals/ijdar/EmbleyKNS16, conf/aaai/NishidaSHM17}, knowledge base construction~\cite{conf/dasfaa/ZhangCCDZ13, conf/kdd/DongGHHLMSSZ14}, knowledge augmentation~\cite{journals/vldb/ElmeleegyMH11, conf/sigmod/SarmaFGHLWXY12, conf/sigmod/SarmaFGHLWXY12, conf/sigmod/YakoutGCC12, conf/www/SekhavatPBM14, conf/sigmod/ChuHCG15, conf/bdc/EberiusBHTAL15, conf/er/BraunschweigTL15}, synonym finding~\cite{journals/pvldb/CafarellaHWWZ08, conf/ijcai/LingHWY13, conf/er/BraunschweigTL15}, improving accessibility~\cite{conf/icdar/PennHLM01, conf/wecwis/WuCY02, conf/das/WangH02, conf/chi/MankoffFT05, conf/hci/OkadaM07}, textual advertising~\cite{conf/cikm/CrestanP10}, data compression~\cite{conf/soda/BuchsbaumCCFM00, conf/icdar/PennHLM01}, or creating linked data~\cite{journals/ijdar/EmbleyKNS16, conf/semweb/KnoblockSFDNSBS17}, just to mention a few common ones.

It is not surprising then that many researchers have worked on a variety of proposals to extract data from tables, which has motivated others to write articles in which they summarise and compare them. \citet{conf/grec/LoprestiN99, conf/iaprgr/LoprestiN99} presented a definition of table, with a focus on how they are encoded and displayed, and motivated the need to extract data from them; they summarised some data extraction techniques, as well as some techniques to integrate the resulting data. \citet{conf/wda/Hurst01} introduced the problem and then reported on some of the challenges regarding locating tables and their cells; he paid special attention to reporting on the evaluation of the proposals and concluded that common evaluation methods are not suitable. \citet{journals/ijdar/ZanibbiBC04} described the extraction tasks as abstract machine-learning procedures in which input documents are first modelled and then mapped onto observations that are transformed prior to performing inference; they analysed many existing proposals according to how they address the steps of the previous procedure; they also highlighted the need for common evaluation methods. \citet{journals/ijdar/SilvaJT06} discussed on what a table is and what makes it different from a diagram; they then listed many proposals to implement the tasks involved in extracting data from tables and compared them using several comparison frameworks; they also criticised common evaluation methods and contributed with some specific purpose evaluation measures. \citet{journals/ijdar/EmbleyHLN06} first discussed on the definition of table and then motivated the need to extract data from them by describing many applications; they listed some proposals to locate tables and their cells, but their emphasis was on the tasks to classify the cells, to group them, and to interpret the tables.

The previous articles focus on the proposals that were published between $1990$ and $2003$.  Unfortunately, there is not a recent article that summarises and compares the proposals that were published later, which motivated us to work on it.  Our focus is on proposals that work on tables that are encoded using HTML because there has been a steady shift towards encoding them using this language~\cite{journals/pvldb/CafarellaHWWZ08, conf/bdc/EberiusBHTAL15, conf/www/LehmbergRMB16}, which provides specific-purpose tags and has become pervasive.  We have analysed $28$ proposals that were published between $2000$ and $2018$, we have defined a vocabulary that homogenises the terminology used in this field, we have used it to summarise the proposals as homogeneously as possible, and we have compared them side by side using several objective characteristics.  We have identified several challenges to which no proposal provides a conclusive solution and also several challenges that have not been addressed sufficiently; addressing them in future shall definitely help produce solutions that increase the range of tables from which data can be extracted correctly. We have also realised that there is not a standardised evaluation method, which hampers the experimental comparison.

The rest of the article is organised as follows: Section~\ref{sec:vocabulary} introduces the vocabulary that we have compiled; Section~\ref{sec:summary-proposals} summarises the proposals that we have analysed using the previous vocabulary; Section~\ref{sec:comparison-proposals} compares them side by side using objective characteristics; finally, Section~\ref{sec:conclusions} concludes the article.


\section{Vocabulary}
\label{sec:vocabulary}

In this section, we have made a point of integrating the many complementary terms that are commonly used in the literature under a common vocabulary.  We first report on the vocabulary that is related to tables themselves and then on the vocabulary that is related to extracting data from them.  We illustrate most of the concepts with a couple of examples.

\subsection{Table-related vocabulary}

Unfortunately, there is not a consensus definition in the literature regarding what a table is.  Many authors focus on the encoding since they define them as whatever one can encode within HTML \code{table} tags~\cite{conf/coling/ChenTT00, conf/icdar/PennHLM01, conf/icdar/PennHLM01, conf/www/Hurst02, conf/widm/YangL02, journals/eaai/KimL05, journals/tkde/JungK06, conf/hci/OkadaM07, journals/pvldb/CafarellaHWWZ08, conf/wsdm/CrestanP11, journals/sigmod/LautertSD13, journals/asc/SonP13, conf/sigmod/ChuHCG15, conf/bdc/EberiusBHTAL15, conf/ksem/WuCWFW16, conf/aisc/LiaoLZL18}, which is a pragmatic approach; a few also refer to the display of data, since they define tables as grids in which data are located in cells in a manner that lines and/or styles ease interpreting them~\cite{conf/icdar/PennHLM01, conf/www/Hurst02, journals/eaai/KimL05, journals/tkde/JungK06, conf/www/GatterbauerBHKP07, conf/ieaaie/FumarolaWBMH11, conf/aaai/NishidaSHM17, journals/ijdar/EmbleyKNS16}.  There is only a proposal that deviates a little from the previous approaches~\cite{journals/ijdar/EmbleyHLN06} since the authors focus on the data model behind the tables, independently from how they are displayed; their proposal, however, works on tables in which data are arranged in grids.

Neither is there a consensus taxonomy of tables.  Most authors differentiate between data tables, which provide data to be extracted, and non-data tables, which are used for layout purposes or to provide utilities.  Many of them make also a difference between listings, forms, matrices, and enumerations~\cite{journals/tkde/JungK06, conf/www/GatterbauerBHKP07, conf/wsdm/CrestanP11, journals/sigmod/LautertSD13, conf/bdc/EberiusBHTAL15, conf/nldb/MilosevicGHN16, conf/ksem/WuCWFW16, conf/aaai/NishidaSHM17}, although the exact terminology used is very diverging; there is also a proposal in which tables are classified according to whether they have headers or not~\cite{journals/ijdar/EmbleyKNS16}.

In the previous discussion, there are three key concepts, namely: encoding, cell, and table, which we define below.

\begin{definition}{Encodings}
An encoding is a specification of how a table must be displayed to a person.  Common encodings include pre-formatted text, images, and mark-ups.  In a table that is encoded using pre-formatted text, the data are arranged in lines, they are aligned to their corresponding columns using blanks, and the cells may be delimited using, for instance, dashes, vertical bars, or tabulators. In a table that is encoded using an image, there is a graphic canvas onto which the data and the lines that delimit the cells, if any, are drawn using bitmaps or vectors.  Contrarily, mark-ups provide a variety of tags that help encode the tables, their cells and, hopefully, additional information that helps interpret them.  There are several mark-up languages available~\cite{journals/software/SierraFF08}, but our focus is on HTML due to its pervasiveness in the Web. HTML provides an array of table-related tags, namely: \code{table}, \code{thead}, \code{tbody}, \code{tfoot}, \code{col}, \code{colgroup}, \code{th}, \code{tr}, \code{td}, and \code{caption}. It is relatively easy to extract data from tables that are encoded using the previous tags.  Unfortunately, real-world tables have a variety of intricacies that hamper the extraction process, namely: some tables are encoded using a subset of table-related tags that hardly help locate them and their cells, which does not help interpret them; other tables are encoded using listing tags (\code{ul}, \code{ol}, \code{dl}, \code{li}, \code{dd}, and \code{dt})~\cite{conf/ijcai/LermanKM01, conf/sigmod/LermanGMK04, journals/vldb/ElmeleegyMH11, conf/sigmod/ChuHCG15}; lately, it is also relatively common to find tables that are encoded using block tags (\code{div} and \code{span}) due to their ability to create responsive layouts~\cite{books/or/Peterson14}; and, generally, speaking, there are many tables that are encoded using a variety of tags that are not actually related to tables, but look like tables when they are displayed~\cite{conf/ieaaie/FumarolaWBMH11, conf/www/GatterbauerBHKP07}.
\end{definition}

\begin{definition}{Cells}
A cell is a box that provides contents to a table.  They can be classified along several axes, namely: \begin{inlinenum} \item According to how they are segmented, cells can be single cells, which occupy exactly one position in the grid of a table, or spanned cells, which occupy more than one position. \item According to whether their contents are complete or not, cells can be classified as single-part cells, whose contents are complete, and multi-part cells, which provide partial contents that must be somewhat merged with the contents of other cells. \item According to their function, they can be classified as meta-data cells, whose contents are labels that help people understand other contents in the table, data cells, whose contents provide the data that must be extracted, decorator cells, which provide irrelevant contents, and context-data cells, which provide captions, notes, or factorised data. \item According to how their contents must be interpreted, cells can be classified as factorised cells, whose contents must be borrowed from adjacent cells, void cells, which are not intended to provide any contents, atomic cells, whose contents cannot be decomposed further, and structured cells, whose contents can be decomposed into a mixture of data and meta-data. \end{inlinenum}
\end{definition}

\begin{definition}{Tables}
A table is a collection of cells that are arranged in rows and columns within a grid, where lines and/or styles are typically used to help people interpret them.  There are cases in which some context data are provided in the text that surrounds a table, i.e., captions, notes, and factorised data.  The cells in a table are typically grouped as follows according to their functions: headers, which are groups of meta-data cells, tuples, which are groups of data cells, and separators, which are groups of decorator cells.  Typically, headers are arranged on the first few rows and/or columns, but we have found some tables in which they are interwoven with tuples for the sake of readability; it was the case of long listings with many tuples, in which it makes sense to repeat the header rows or columns every few tuples, or wide listings/forms with many headers, in which it makes sense to split the header rows or columns to narrow them. Data tables can be broadly classified as follows: \begin{inlinenum} \item listings, in which the headers, if any, occupy either the first few rows or columns and the tuples are arranged in the remaining rows or columns, respectively; \item forms, in which the headers, if any, occupy either the first few rows or columns and there is a single tuple that is arranged row- or column-wise, respectively; \item matrices, in which the headers occupy both the first few rows and columns, and all of the data cells constitute a single tuple; and \item enumerations, in which there are no headers and each individual cell can be considered a tuple. \end{inlinenum}  According to \citet{conf/wsdm/CrestanP11}, this taxonomy covers roughly $98\%$ of the data tables in their $8.2$-billion table repository; the authors mention that it is arguable that the remaining $2\%$ tables can be considered actual data tables or that they are frequent enough to be representative.
\end{definition}

\begin{figure}
	\includegraphics[width=40em]{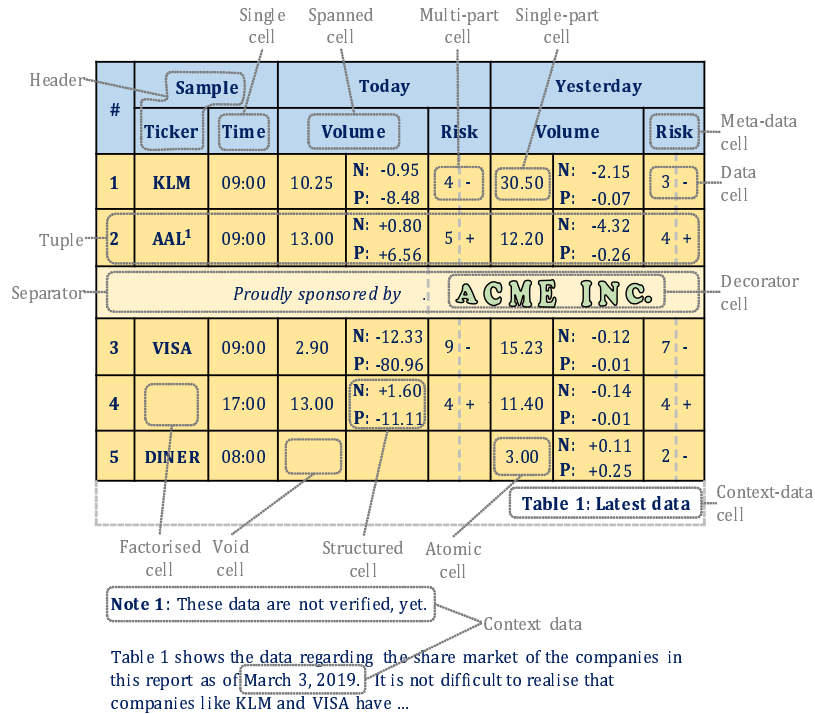}
    \caption{A sample table.}
	\label{fig:sample-table}
\end{figure}

\begin{example}
Figure~\ref{fig:sample-table} shows a horizontal listing taken from a document on a share market.  The black, solid lines help delimit the boundaries of the cells in the grid; the greyed, dashed lines represent the boundaries of a few cells that exists in the encoding of the table, but are not visible to the reader because they are used for layout purposes.  Most of the data are displayed on an $9 \times 11$ grid, but there are also some context data in the surrounding text.

Cells like \qcode{Time} are single because they occupy exactly one position in the grid; on the contrary, cell \qcode{Volume} is spanned because it occupies two positions in the grid.  The meta-data cells occupy the first two rows; for instance \qcode{Risk} is one such meta-data cell.  Contrarily, cell \qcode{3-} below is a data cell and cell \qcode{Acme Inc.} is a decorator cell. The caption of the table is displayed within a bottom cell that spans the whole table, which is considered a context-data cell; realise that there are additional context data: there is a note regarding cell \qcode{AAL} and there is a factorised datum regarding the date of the report, which complements the times provided in some cells. Cells like \qcode{30.50} are single-part cells because their contents are complete; on the contrary, cells like \qcode{4-} are multi-part cells because it is necessary to merge the contents of two cells so that the contents of the resulting cell are complete. A cell like \qcode{3.00} is atomic since its contents cannot be decomposed further; contrarily, a cell like \qcode{N: +1.60 P: -11.04} is a structured cell because it provides both meta-data and data, which means that it can be decomposed further. The empty cell below cell \qcode{VISA} very likely factorises the ticker since there are two tuples regarding this company at 09:00 and 17:00, respectively; contrarily, the four empty cells on the right of cell \qcode{08:00} are very likely void cells that indicate that no data are available.

The table has seven headers, namely: \qcode{\#}, \qcode{Sample/Ticker}, \qcode{Sample/Time}, \qcode{Today/Volume}, \qcode{Today/Risk}, \qcode{Yesterday/\-Volume}, and \qcode{Yesterday/Risk}.  It provides five tuples, the first of which is (\qcode{1}, \qcode{KML}, \qcode{09:00}, \qcode{10.25}, \qcode{N: -0.95 P: -8.48}, \qcode{4-}, \qcode{30.50}, \qcode{N: -2.15 P: -0.07}, \qcode{3-}).  It has also a separator at the fifth row, which shows an advertisement.
\end{example}

\subsection{Data-extraction vocabulary}

In our context, data extraction refers to a process that transforms the tables in an input document into record sets.  A record is a data structure in which the individual data in a tuple are endowed with semantics by means of descriptors that are computed from the meta-data provided by the corresponding table; in cases in which the table does not provide enough meta-data, the descriptors must be generated artificially.

\citet{journals/ijdar/SilvaJT06} did a good job at identifying the tasks of which the data-extraction process is composed, namely: location, segmentation, functional analysis, structural analysis, and interpretation.  Note, however, that their focus was on tables that are encoded using pre-formatted text or images, which means that they need not make tables that provide data apart from tables that are intended for layout purposes or to provide utilities. The latter are very common in nowadays Web, which motivated \citet{journals/pvldb/CafarellaHWWZ08}, for instance, to introduce a task to discriminate data tables from non-data tables.

Before feeding the record sets returned by data extraction into a particular application, it is commonly necessary to perform some of the following integration tasks: semantisation~\cite{conf/semweb/MulwadFSJ10a, journals/pvldb/VenetisHMPSWMW11, conf/sigmod/ZhangC13, conf/edbt/RitzeB17, conf/kesw/GalkinMA15, journals/ws/TaheriyanKSA16, conf/www/RenWHQVJAH17}, which either maps the descriptors onto the terminology box of a particular ontology or the tuples onto its assertion box~\cite{conf/semweb/EfthymiouHRC17}; union~\cite{conf/icde/FanLOTZ14}, which merges record sets that provide similar data; finding primary keys~\cite{journals/tods/TschirschnitzPN17}, which determines which components of the tuples identify them as univocally as possible; record linkage~\cite{books/springer/Christen12, conf/hais/CimminoC18, conf/bis/CimminoC18}, which finds different records that refer to the same actual entities; augmentation~\cite{conf/sigmod/YakoutGCC12, conf/IEEEwisa/QiWW17, conf/webdb/CannaviccioABM18}, which joins record sets on the same topic to complete the information that they provide individually; and cleaning~\cite{conf/sigmod/KhayyatIJMOPQ0Y15, conf/bigdata/TalebDS15, conf/sigmod/ChuMIOP0Y15}, which fixes data.  Note that the integration tasks are orthogonal to data extraction because they are independent from the source of the record sets, which is the reason why they fall out of the scope of this article.

In the previous discussion, there are three key concepts: record set, extraction task, and data extraction, which we define below.

\begin{definition}{Record set}
A record set is a collection of records.  A record is a map that associates a set of descriptors to each of the components of a tuple.  A descriptor is a structured label that endows the components of a tuple with the semantics provided by the meta-data in the corresponding headers or structured cells; if not enough meta-data are available, then descriptors must be generated artificially. We make three types of descriptors apart, namely: simple descriptors, which correspond to the contents of a single meta-data cell, field descriptors, which correspond to the contents of several adjacent meta-data cells, and artificial descriptors, which are used when not enough meta-data are available.  In listings and forms, every component of the tuples has one associated descriptor; in matrices, they have two associated descriptors; in enumerations, the descriptors must be created from the meta-data in the cells, if any; in other cases, they must be generated artificially.
\end{definition}

\begin{definition}{Extraction tasks}
The tasks involved in extracting data from a table are the following~\cite{journals/ijdar/SilvaJT06}: \begin{inlinenum} \item location, which searches the input document for the excerpts in which tables are encoded and returns them; \item segmentation, which searches for the cells of which a table is composed; \item discrimination, which classifies a table as either a data table or a non-data table, but further sub-classification is possible; \item functional analysis, which classifies the cells according to their functions; \item structural analysis, which groups cells into at least headers and tuples; and \item interpretation, which produces record sets building on the results of the previous tasks. \end{inlinenum}
\end{definition}

\begin{definition}{Data extraction}
Data extraction refers to a process that organises the extraction tasks into a pipeline so that they can achieve their goal.  \citet{journals/ijdar/ZanibbiBC04} and \citet{journals/ijdar/SilvaJT06} reported on the many common inter-dependencies amongst the extraction tasks.
\end{definition}

\begin{figure}
	\includegraphics[width=35em]{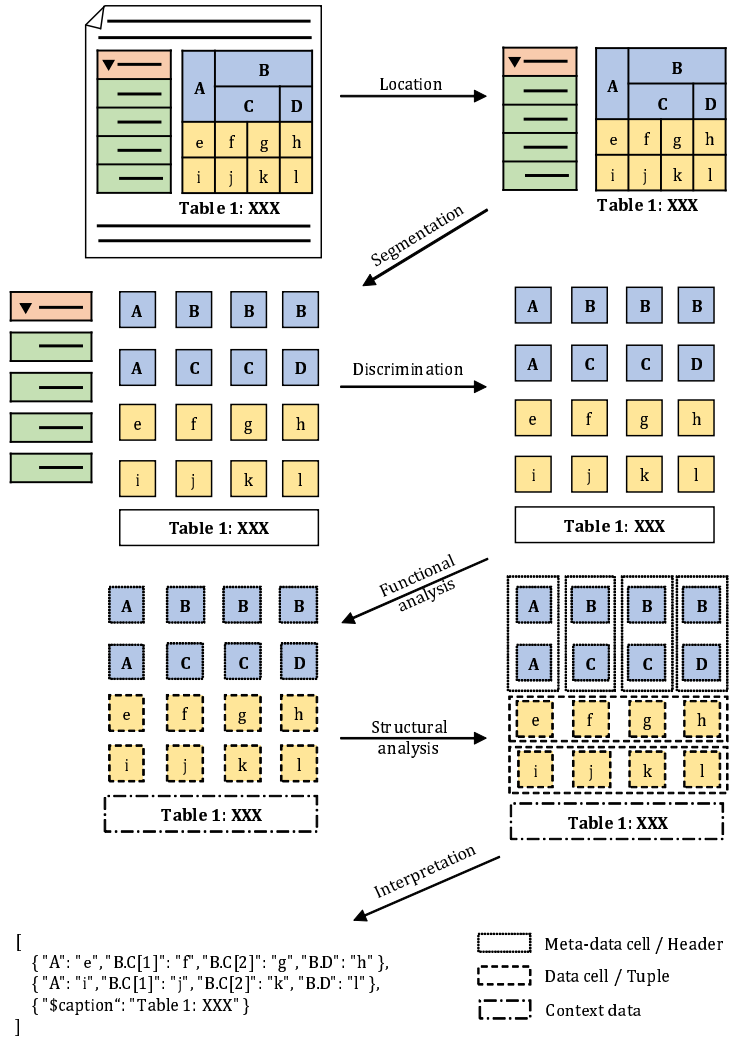}
    \caption{A sample data extraction process.}
	\label{fig:sample-data-extraction-process}
\end{figure}

\begin{example}
Figure~\ref{fig:sample-data-extraction-process} illustrates a sample data extraction process in which we have organised the tasks into a sequential pipeline.

The location task finds two excerpts in the input document that seem to have tables; the segmentation task is responsible for finding the individual cells of which the tables are composed, plus the context data that is associated with them; the discrimination task makes a difference between the table on the left, which seems to be a menu that does not provide any data, and the table on the right, which seems to be a table that provides data; the functional analysis task makes meta-data cells apart from data cells; the structural analysis task groups the meta-data cells into four headers and the data cells into two tuples; finally, the interpretation task produces a record set with three records.

Regarding the descriptors, we illustrate them using the usual field-access notation for simple and field descriptors and the usual array-access notation for artificial descriptors.  For instance, header \qcode{A/A} results in a simple descriptor of the form \qcode{A} because both cells were actually a vertically-spanned cell in the original table.  On the contrary, header \qcode{B/C} results in a field descriptor of the form \qcode{B.C} in which it is clear that whatever \qcode{C} represents is subordinated to whatever \qcode{B} represents; note that this descriptor is ambiguous since there are two columns of the table with the same header.  In such cases, the table does not provide enough meta-data and the columns must be made apart by means of artificial descriptors, that is \qcode{B.C[1]} and \qcode{B.C[2]}.  Obviously, header \qcode{B/D} results in a field descriptor of the form \qcode{B.D}.

The records extracted are the following: \code{\{"A": "e", "B.C[1]": "f", "B.C[2]": "g", "B.D": "h"\}}, \code{\{"A": "i", "B.C[1]": "j", "B.C[2]": "k", "B.D": "l"\}}, and \code{\{"\$caption": "Table 1: XXX"\}}.  Realise that the last record uses a special simple descriptor to indicate that corresponding datum is the caption of the table.
\end{example}


\section{Summary of proposals}
\label{sec:summary-proposals}

\begin{table*}
    \begin{adjustbox}{angle=90}
    	\includegraphics[width=45em]{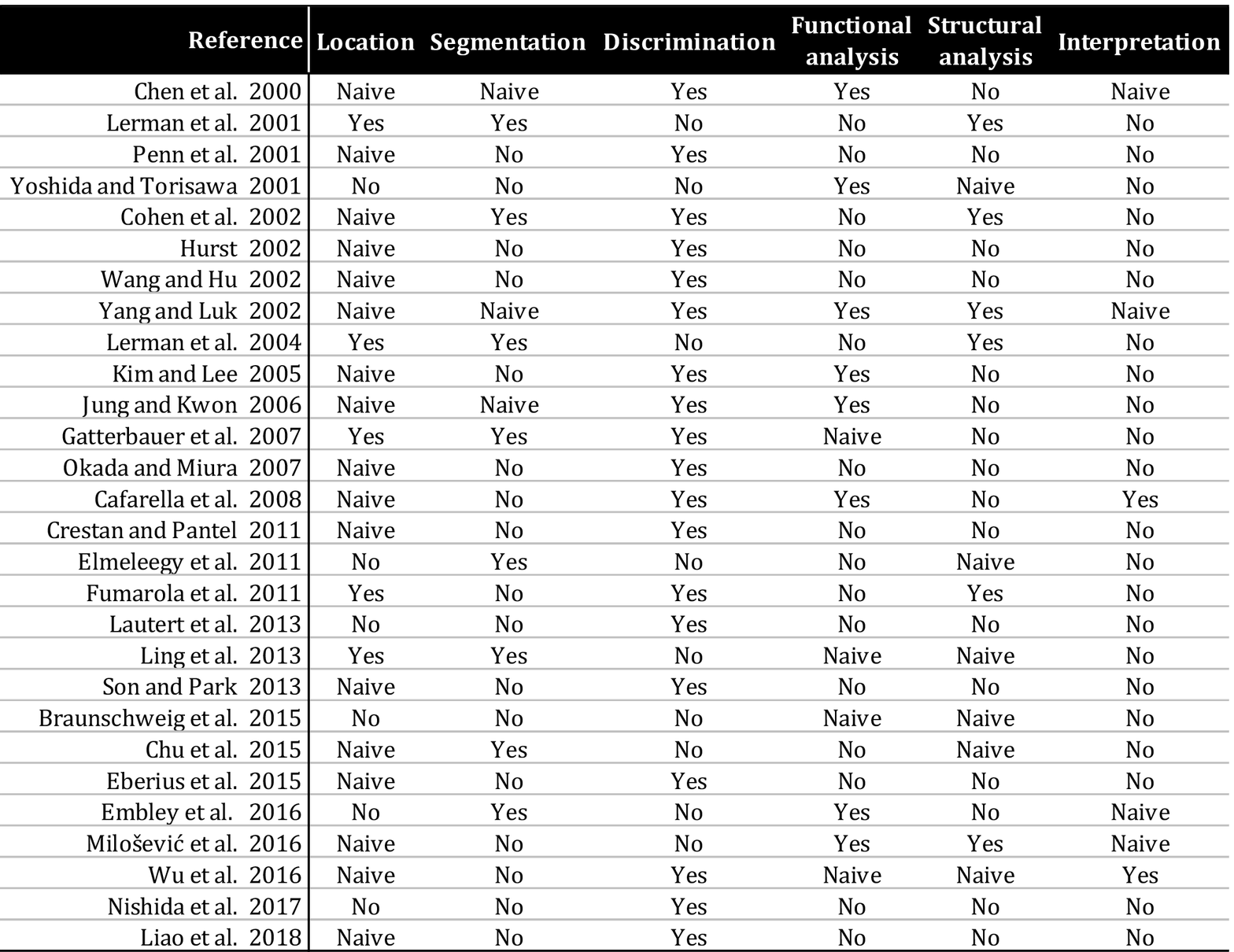}
    \end{adjustbox}
    \caption{Tasks addressed by each proposal.}
	\label{tab:tasks-addressed-proposals}
\end{table*}

In this section, we summarise many proposals that have not been surveyed previously, cf.~Table~\ref{tab:tasks-addressed-proposals}.  Note that our summaries are intended to provide an overall picture using the vocabulary that we have presented previously; please, refer to the original articles for the full descriptions.

\subsection{Location}

\citet{conf/wda/YoshidaT01}, \citet{journals/vldb/ElmeleegyMH11}, \citet{journals/sigmod/LautertSD13}, \citet{conf/er/BraunschweigTL15}, \citet{journals/ijdar/EmbleyKNS16}, and \citet{conf/aaai/NishidaSHM17} did not pay attention to the location task. \citet{conf/coling/ChenTT00}, \citet{conf/www/CohenHJ02}, \citet{conf/www/Hurst02}, \citet{conf/widm/YangL02}, \citet{journals/eaai/KimL05}, \citet{journals/tkde/JungK06}, \citet{conf/hci/OkadaM07}, \citet{journals/pvldb/CafarellaHWWZ08}, \citet{journals/asc/SonP13}, \citet{conf/sigmod/ChuHCG15}, \citet{conf/bdc/EberiusBHTAL15}, \citet{conf/nldb/MilosevicGHN16}, \citet{conf/ksem/WuCWFW16}, and \citet{conf/aisc/LiaoLZL18} reported on naive approaches that consisted in extracting every HTML excerpt with a \code{table} tag; \citet{conf/icdar/PennHLM01}, \citet{conf/das/WangH02}, and \citet{conf/wsdm/CrestanP11} followed the same approach but discarded tables with nested tables. The other proposals provide more sophisticated approaches.

\citet{conf/ijcai/LermanKM01, conf/sigmod/LermanGMK04} focused on tables that are encoded using listing tags.  Their proposal works as follows: \begin{inlinenum} \item first, the input documents are tokenised and the tokens are assigned to lexical types; \item then, the smallest input document is taken as a base template; \item the remaining documents are then iteratively compared to the template in order to make the sequences of tokens that appear exactly once apart from the others; \item finally, the excerpts of the document that have the largest repetitive sequence of tokens are returned. \end{inlinenum} The authors did not evaluate their procedure in isolation, but their complete system.

\citet{conf/www/GatterbauerBHKP07} presented a visual approach that analyses the bounding boxes used to display the elements of a document in an attempt to identify tables, lists, and so-called aligned graphics that represent tabular structures.  Their proposal works as follows: \begin{inlinenum} \item first, they apply some heuristics to locate elements in the DOM tree that are likely to be a part of a table, a list, or an aligned graphic; the authors mention that they have compiled a collection of over twenty such heuristics, but they document only twelve of them in their paper; \item then, they apply an algorithm that searches for so-called frames, which are collections of elements that are rendered so that they form a box; \item then, the frames are expanded to four orthogonal directions by finding elements whose bounding boxes are near to each other; \item finally, the excerpts that correspond to the extended frames are returned. \end{inlinenum}  The authors  evaluated their proposal on $493$ tables from their own repository plus $19$ additional tables from \cites{conf/das/WangH02} repository.

\citet{conf/ieaaie/FumarolaWBMH11} also presented a visual approach.  Their proposal works as follows: \begin{inlinenum} \item it first creates a bounding box that encloses the whole input document; \item it then iterates recursively and creates a bounding box for every element in that document; \item next, it analyses the positions of the inner bounding boxes and finds those that are laid out in a row- or a column-wise manner; \item then, the corresponding excerpts are returned. \end{inlinenum} The authors did not evaluate their procedure in isolation, but their complete system.

\citet{conf/ijcai/LingHWY13} presented a proposal whose focus is on locating context-data cells. It works as follows: \begin{inlinenum} \item first, it locates the elements in the input document that have a \code{table} tag; \item then, it extracts some context data from the \code{title} tag; \item next, it segments the text around the tables and aligns the resulting segments using a multiple string alignment algorithm; \item finally, the segments that are repetitive enough are considered context-data cells. \end{inlinenum}  The authors did not evaluate their procedure in isolation, but their whole system.

\subsection{Segmentation}

\citet{conf/icdar/PennHLM01}, \citet{conf/wda/YoshidaT01}, \citet{conf/www/Hurst02}, \citet{conf/das/WangH02}, \citet{journals/eaai/KimL05}, \citet{conf/hci/OkadaM07}, \citet{journals/pvldb/CafarellaHWWZ08}, \citet{conf/wsdm/CrestanP11}, \citet{conf/ieaaie/FumarolaWBMH11}, \citet{journals/sigmod/LautertSD13}, \citet{journals/asc/SonP13}, \citet{conf/er/BraunschweigTL15}, \citet{conf/bdc/EberiusBHTAL15}, \citet{conf/nldb/MilosevicGHN16}, \citet{conf/ksem/WuCWFW16}, \citet{conf/aaai/NishidaSHM17}, and \citet{conf/aisc/LiaoLZL18} did not report on any proposals to implement this task. \citet{conf/coling/ChenTT00}, \citet{conf/widm/YangL02}, \citet{journals/tkde/JungK06} relied on a naive approach that searches for the cells using specific tags. The other proposals provide more sophisticated approaches.

\citet{conf/ijcai/LermanKM01} focused on tables that are encoded using listing tags; implicitly, they assumed that tuples are shown in a row-wise manner. Prior to segmentation, the authors applied a document alignment method to detect the template of the documents and their repetitive segments, which are very likely to contain the lists. Once the lists are located, their proposal works as follows: \begin{inlinenum} \item first, the segments are grouped according to their separators; \item then, DataPro is invoked on the previous groups to learn patterns that characterise their data; \item then, for each segment in each group, it computes binary features that indicate whether it matches the previous patterns or not; \item next, the AutoClass clustering algorithm is invoked to learn the optimal number of clusters and to learn a set of rules that assign new segments to the most similar cluster; \item finally, the data in each cluster is assumed to be a column of the corresponding table, which facilitates identifying the cells in a row-wise manner. \end{inlinenum} The evaluation was performed on the tables from a repository with $50$ documents that were taken from $14$ different sources.  The authors did not evaluate their procedure in isolation, but their complete system.

\citet{conf/www/CohenHJ02} relied on some transformations that help normalise tables before they are segmented. Their proposal works as follows: \begin{inlinenum} \item the HTML structure is cleaned up using HTML Tidy and the extra cells generated by this tool are removed; \item structured cells are divided into multiple atomic cells by splitting inner tables, paragraphs, or pre-formatted text; \item spanned cells are split into several cells unless this results in more cells than the height or the width of the table. \end{inlinenum} The authors did not evaluate their procedure in isolation, but their complete system.

\citet{conf/sigmod/LermanGMK04} segmented tables by learning a probabilistic model from the repetitive segments in which they decompose tables that are encoded using listing tags; they assumed that tuples are shown in a row-wise manner.  Their proposal works as follows: \begin{inlinenum} \item lists are split into columns according to candidate separators, which can be tags or punctuation symbols; \item some content features are then computed on each column and their siblings; \item then, an inference algorithm learns a probabilistic model from the previous features; \item the parameters are then used to find the best column assignment for a segment, which is the one that maximises the probability of the features observed given the model. \end{inlinenum} Their evaluation was performed on the tables from a repository with $283$ tables from $12$ web sites on book sellers, property taxes, white reports, and corrections. They also experimented with a constrain satisfaction approach that was less accurate.

\citet{conf/www/GatterbauerBHKP07} presented a proposal that requires to identify the spatial relationships between the individual cells of a table. It works as follows: \begin{inlinenum} \item it computes the boxes that represent the elements of an input document, taking into account their contents area, padding, border, and margin areas according to the CSS2 visual formatting model; \item it then overlaps a grid that helps identify each box by means of the co-ordinates of its upper-left corner and its lower-right corner; \item then, it aligns the boxes according to their horizontal and vertical projections; \item next, an adjacency relation is computed according to how distant the cells are; \item finally, some cells are selected and a recurrent expansion algorithm is invoked in an attempt to explore the adjacency relation to find their neighbours. \end{inlinenum} The authors evaluated their proposal on the tables provided by a repository with $1\,537$ documents that were retrieved from search engines, from \cites{conf/das/WangH02} repository, or written by the authors. The authors did not evaluate their procedure in isolation, but their complete system.

\citet{journals/vldb/ElmeleegyMH11} tried to find columns by checking how similar the cells in a table are. The similarity is analysed by means of their data types and delimiters. The authors used two resources, a large-scale language model, which helps know sentences that should not be split because they have previously occurred within a cell, and a corpus of tables, which helps identify data that appear in the same column in other tables. Their proposal works as follows: \begin{inlinenum} \item each row is split into a (possibly different) number of columns using two scoring functions, namely: a field quality score, which measures the quality of an individual column candidate, and a field-to-field consistency score, which measures the likelihood that two column candidates are actually the same column; \item then, it sets the number of columns to the most frequent one; \item padding columns are added to rows that have less columns than expected and some columns are merged otherwise; \item finally, the segmentation of cells is refined by checking the consistency amongst the  cells in a per-column basis; \item if the consistency check fails, the procedure is re-launched. \end{inlinenum} Their evaluation was performed on $20$ tables from $20$ different domains plus $100$ additional tables that were randomly sampled from the Web.

\citet{conf/ijcai/LingHWY13} assumed that tables can be segmented building on their \code{td} tags; their key contribution was regarding how to find context data.  Their proposal works as follows: \begin{inlinenum} \item it first uses a number of heuristics to generate candidate context data, namely: tokens in between some punctuation marks, the longest common sub-sequences, pieces of text that can be wikified~\cite{conf/acl/RatinovRDA11}, and pieces of text that vary from document to document but are located at the same position; \item then, the previous context data are added to the original table as additional columns; \item finally, a pairwise adaptation of the Multiple Sequence Alignment algorithm is used to segment the context data. \end{inlinenum} The evaluation was performed on $20\,000$ tables that were picked from a repository with $130$ million tables from $10$ different web sites.

\citet{conf/sigmod/ChuHCG15} also focused on finding the columns of a data table. Their proposal works as follows: \begin{inlinenum} \item each row is tokenised using a set of user-defined delimiters; \item then candidate columns are generated using two approaches: a seed tuple is provided and the system discards segmentations that are very different; a custom pruning procedure that borrows some ideas from the well-known A* procedure is also used; \item it then measures the similarity of each column using lexical and semantic similarity functions that are averaged (the former computes the difference regarding the number of tokens, characters, and pattern-based types; the latter computes the point-wise mutual information function); \item the process is repeated until a segmentation that maximises similarity is found. \end{inlinenum}  Their evaluation was performed on $100$ million data tables that were transformed into lists; they used $20$ additional tables encoded as lists from five different domains.

\citet{journals/ijdar/EmbleyKNS16} presented a proposal that works as follows: \begin{inlinenum} \item the input documents are transformed into a representation that preserves the contents only; \item spanned cells are split and their contents are copied verbatim to the resulting cells; \item then, every row with more than two empty cells is considered to provide context data; \item finally, the right-most bottom non-empty cell is considered to be the last cell in the table.  (Note that they can work on tables that come from spreadsheets, in which it is not uncommon to find empty cells that are not actually part of any tables.) \end{inlinenum} The authors did not evaluate their procedure in isolation, but their complete system.

\subsection{Discrimination}

\citet{conf/ijcai/LermanKM01}, \citet{conf/wda/YoshidaT01}, \citet{conf/sigmod/LermanGMK04}, \citet{journals/vldb/ElmeleegyMH11}, \citet{conf/ijcai/LingHWY13}, \citet{conf/er/BraunschweigTL15}, \citet{conf/sigmod/ChuHCG15}, \citet{journals/ijdar/EmbleyKNS16}, and \citet{conf/nldb/MilosevicGHN16} did not pay attention to the discrimination task. The other proposals provide sophisticated approaches.

\citet{conf/coling/ChenTT00} devised a proposal to discriminate tables by means of heuristics. It works as follows: \begin{inlinenum} \item a cell similarity measure is computed by combining string similarity, named entity similarity, and number similarity functions; \item then, the tables whose cells do not exceed a threshold regarding the number of similar neighbour cells are discarded; \item finally, tables with less than two cells or tables with many links, forms, or figures, are also discarded. \end{inlinenum} The evaluation was performed on $3\,218$ tables from their own repository with documents on airlines from the Chinese Yahoo! site.

\citet{conf/icdar/PennHLM01} also devised a heuristic-based approach. Their proposal works as follows: \begin{inlinenum} \item tables that do not have multiple rows and columns are discarded; \item tables whose cells have more than one non-text-formatting tag are also discarded; \item finally, tables whose cells have more than a user-defined number of words are also discarded. \end{inlinenum} The authors also mentioned that a desirable feature is to have syntactic and semantic similarity into account, but they did not explore this idea. They experimented with an unspecified number of tables from their own repository with documents from $75$ sites on news, television, radio, and companies.

\citet{conf/www/CohenHJ02} devised a proposal that builds on machine learning a classifier. It works as follows: \begin{inlinenum} \item some structural and content features are computed from a learning set with tables that are pre-classified as either data tables or non-data tables; \item then, several classifiers are machine-learnt and evaluated; \item the classifier that achieves the best effectiveness is selected to implement the discrimination task. \end{inlinenum} The authors experimented with Multinomial Naive Bayes, Maximum Entropy, Winnow, and a decision tree learner that was based on C4.5; their conclusion was that the best results were achieved using Winnow.  They evaluated their proposal using a $5$-trial approach on $339$ tables from their own repository; in each trial, $75\%$ of the tables were used for learning and the remaining $25\%$ for evaluation purposes.

\citet{conf/www/Hurst02} presented another machine-learning approach in which he also took visual features into account. He performed his evaluation on $89$ data tables and $250$ non-data tables from his own repository; they were randomly grouped into five sets from which $25\%$ of the tables were selected for learning purposes and $75\%$ for evaluation purposes. The results confirmed that Naive Bayes achieved the best results when the whole set of features was used, whereas Winnow worked better when only geometric features were used.

\citet{conf/das/WangH02} devised another machine-learning proposal that relies on structural and content features that are used to feed a custom decision tree learner; some of the features need to be transformed into real values using Naive Bayes or $k$-NN.  The content features rely on the words found in the input documents, which requires a large learning set so as to minimise the chances that a classifier is applied to a document with a word that was not in the learning set. The evaluation was performed using $9$-fold cross evaluation on $11\,477$ tables from their own repository with documents from Google's directories.

\citet{conf/widm/YangL02} reported on another heuristic-based method. Their proposal works as follows: \begin{inlinenum} \item tables that have \code{th} tags are considered data tables; \item tables that do not only contain links, forms, or images are also considered data tables; \item meta-data and data cells are then located using some user-defined patterns; \item tables that do not have both meta-data and data cells are discarded. \end{inlinenum} They evaluated their method on $1\,927$ tables from their own repository, which was assembled with random documents from the Web.

\citet{journals/eaai/KimL05} used heuristics and an algorithm to check how similar the cells are. Their proposal works as follows: \begin{inlinenum} \item tables are considered data tables if they contain \code{caption} or \code{th} tags and there are \code{td} tags at the right or the bottom sides; \item they are discarded if they have a single cell, if they have nested tables, or if they seem to have meta-data cells only; \item if they have too many links, images, or empty cells, then they are also discarded; \item then, it checks that the cells selected previously are consistent using some user-defined patterns; \item if the degree of similarity per row or column does not exceed a pre-defined threshold, then the corresponding table is discarded. \end{inlinenum}  The evaluation was performed on $11\,477$ tables from \cites{conf/das/WangH02} repository.

\citet{journals/tkde/JungK06} presented a machine-learning proposal. It works as follows: \begin{inlinenum} \item it first removes empty rows and columns, splits spanned cells by duplicating their contents, and discards tables with only one cell; \item then, it computes many structural, visual, and content features of the table to find out if it has meta-data cells, in which case the table is assumed to have data; \item finally, a C4.5 learner is fed with the input features and the classified tables. \end{inlinenum}  The evaluation was performed using $10$-fold cross evaluation on $10\,000$ tables from their own repository plus some tables from \cites{conf/das/WangH02} repository.

\citet{conf/www/GatterbauerBHKP07} reported on an approach that identifies tables using some display heuristics. Their proposal works as follows: \begin{inlinenum} \item elements with \code{td}, \code{th}, and \code{div} tags are considered candidate tables; \item it tries to identify frames that rely on those elements, which are assumed to be tables; \item overlapping tables are discarded; \item tables are also discarded if, after removing separator columns and rows, they have less than three rows, a single cell is more than $40\%$ the total size of the table, or they contain cells with more than $20$ words. \end{inlinenum}  The evaluation was performed on $493$ tables from their own repository.

\citet{conf/hci/OkadaM07} devised another machine-learning approach that requires to binarise discrete features before feeding them into an ID3 learner.  The evaluation was performed using $10$-fold cross evaluation on $100$ data tables and $100$ non-data tables from their own repository.

\citet{journals/pvldb/CafarellaHWWZ08} proposed another machine-learning approach. Their proposal works as follows: \begin{inlinenum} \item it considers tables that have at least four cells, are not embedded in HTML forms, and are not calendars; \item the tables that meet the previous criteria are classified as either data or non-data tables by a person; \item then, a statistical classifier is machine-learnt from a dataset that vectorises the previous tables using both structural and content features that are intended to measure how consistent the cells are.  \end{inlinenum}  They evaluated their proposal using $5$-fold cross evaluation over several thousand tables from their own repository.

\citet{conf/wsdm/CrestanP11} also presented a machine-learning proposal. It works as follows: \begin{inlinenum} \item tables that have less than four cells or have cells with more than $100$ characters are discarded; \item next, some structural and content features are computed; \item then, a Gradient Boosted Decision Tree classification model is machine-learnt. \end{inlinenum} The evaluation was conducted on $5\,000$ tables from their own repository by performing $20$-fold cross evaluation without overlapping.

\citet{conf/ieaaie/FumarolaWBMH11} proposed a heuristic-based approach. Their proposal works as follows: \begin{inlinenum} \item it groups the elements whose bounding boxes are arranged in a grid; \item it then computes their similarity by comparing their DOM trees; \item next, it computes the number of nodes in each group; \item if the similarity in a group is above a user-defined threshold and the difference in the number of nodes is below another user-defined threshold, then it is considered a data table. \end{inlinenum} The evaluation was performed on $224$ tables that were gathered from \cites{conf/www/GatterbauerBHKP07} repository.

\citet{journals/sigmod/LautertSD13} devised a machine-learning proposal that builds on neural networks. It works as follows: \begin{inlinenum} \item it computes some structural, visual, and content features; \item then, it uses them to machine-learn a perceptron with one hidden layer and resilient propagation; \item it has one output neuron per type of data table, which is encoded using a score in range $[0.00 \upto 1.00]$; the classification is performed in two steps, namely: the first one uses $25$ features to classify the tables into the corresponding types and the second step uses the previous $25$ features plus the type of table output by the previous classifier. \end{inlinenum} The evaluation was performed on a repository with $342\,795$ tables that were gathered randomly.

\citet{journals/asc/SonP13} also tried a machine-learning approach.  Their proposal works as follows: \begin{inlinenum} \item it selects every DOM node with tag \code{table} and their corresponding parents; \item the features described by \citet{conf/das/WangH02} are then computed to create a learning set; \item finally, an SVM classifier is machine-learnt using a kernel that works with structural features plus a kernel that works with content features; the structure kernel is based on two other kernels, one of which works on the table nodes and the other on the corresponding parent nodes.  \end{inlinenum}  The authors performed $10$-fold cross evaluation on a subset of $11\,477$ tables from \cites{conf/das/WangH02} repository; roughly $89\%$ of the tables were used for learning purposes and roughly $11\%$ were used for evaluation purposes.

\citet{conf/bdc/EberiusBHTAL15} devised a proposal that builds on machine learning a classifier. It works as follows: \begin{inlinenum} \item some heuristics are applied to filter most non-data tables out, namely: tables with less than two rows or columns, tables with an invalid HTML structure, and tables that cannot be displayed correctly; \item some structural and content features are then computed regarding the tables and some of their subregions in order to compute local features; \item two alternatives are now tried: learning one classifier for every table type or using one classifier to discriminate between data and non-data tables and an additional classifier to classify some kinds of data tables; \item several classifiers are machine-learnt and evaluated, namely: CART, C4.5, SVM, and Random Forest; \item the classifier that achieves the best effectiveness is selected to implement the discrimination task. \end{inlinenum} They evaluated their proposal on a repository with $24\,654$ tables from the October 2014 Common Crawl.  According to their experience, the best results were achieved with Random Forest.

\citet{conf/ksem/WuCWFW16} provided a method to cluster tables that are similar according to their structure. Their proposal works as follows: \begin{inlinenum} \item for every two tables, it computes the set of paths that corresponds to \code{caption}, \code{td}, and \code{th} tags; \item then, the similarity between the paths of every two tables is computed; \item then, tables are clustered according to their local density plus the previous similarities; \item now, for each cluster, clustering is performed again building on the paths that lead to elements with tags \code{li}, \code{span}, or \code{div}; \item finally, a so-called artificial judgment method is used to decide on the class of each cluster. \end{inlinenum} The authors used a repository with $5\,000$ tables from the Wikipedia to evaluate their system, but no results were provided regarding this task.

\citet{conf/aaai/NishidaSHM17} devised a proposal that analyses a subset of cells at the top-left corner of a table using a deep neural network. It works as follows: \begin{inlinenum} \item for each \code{td} or \code{th} tag, an embedding is generated by tokenising words, tags, and row and column indexes; \item each token is encoded as a one-hot vector; \item an LSTM with an attention mechanism is then used to obtain a semantic representation of each cell; \item a convolutional neural network is then connected to three residual units and applied to vectorise the input table; \item finally, a classification layer is used. \end{inlinenum} The authors learnt the network using $3\,567$ tables from $200$ web sites, and evaluated the results on $60\,678$ tables from $300$ web sites; the documents were selected from the April 2016 Common Crawl.  They also experimented with an ensemble of five neural networks, which attained the best results.

\citet{conf/aisc/LiaoLZL18} presented a heuristic-based approach that takes into account the existence of nested data tables. It works as follows:  \begin{inlinenum} \item tables with a \code{th} or \code{caption} tag are considered data tables; \item tables with a large number of pictures, frames, forms, or script tags are discarded; \item tables with a small number of elements or many empty cells are discarded, too; \item tables with too many homogeneous contents in their rows are considered incomplete data tables, which must be stitched to other sibling tables to create a complete data table. \end{inlinenum} They evaluated their method on $226$ tables from $50$ different sites.

\subsection{Functional analysis}

\citet{conf/ijcai/LermanKM01}, \citet{conf/icdar/PennHLM01}, \citet{conf/www/CohenHJ02}, \citet{conf/www/Hurst02}, \citet{conf/das/WangH02}, \citet{conf/sigmod/LermanGMK04}, \citet{conf/hci/OkadaM07}, \citet{conf/wsdm/CrestanP11}, \citet{journals/vldb/ElmeleegyMH11}, \citet{conf/ieaaie/FumarolaWBMH11}, \citet{journals/sigmod/LautertSD13}, \citet{journals/asc/SonP13}, \citet{conf/sigmod/ChuHCG15}, \citet{conf/bdc/EberiusBHTAL15}, \citet{conf/aaai/NishidaSHM17}, and \citet{conf/aisc/LiaoLZL18} did not report on any proposals to implement the functional analysis task.  \citet{conf/www/GatterbauerBHKP07} presented a naive approach that matches the structure of a table to a number of pre-defined structures in which it is also relatively easy to find the meta-data cells. \citet{conf/ijcai/LingHWY13} and \citet{conf/ksem/WuCWFW16} assumed that meta-data cells can be easily located by searching for \code{th} tags. \citet{conf/er/BraunschweigTL15} also presented a naive solution since they assumed that meta-data cells are located on the first row. The other proposals provide more sophisticated approaches.

\citet{conf/coling/ChenTT00} devised a proposal that is based on row/column similarity.  It works as follows: \begin{inlinenum} \item it first divides the input table into blocks using the spanned cells as boundaries; \item it them compares how similar the last row/column in each block is to the previous ones using string, named-entity, and number similarity functions; \item then the right-most and/or bottom-most rows/columns that are similar to the last row/column are considered to contain data cells and the others are considered to contain meta-data cells. \end{inlinenum} The evaluation was not performed on this task, but on their whole system.

\citet{conf/wda/YoshidaT01} suggested using ontologies.  Their proposal works as follows: \begin{inlinenum} \item for each cell in a table, it computes the ratio of times that its content is recorded in the ontology; \item these ratios are then used to feed the Expectation-Maximisation algorithm in order to learn a classifier that makes a few subtypes of listings apart; \item once the exact type of listing is clear, identifying meta-data cells is relatively easy and the rest of cells are assumed to be data cells. \end{inlinenum} (Note that the authors assume that the input tables are data tables, which is the reason why this cannot be considered a discrimination proposal.)  They evaluated their proposal on $175$ tables that were randomly sampled from a repository with $35\,232$ tables.

\citet{conf/widm/YangL02} applied some heuristics to differentiate rows with meta-data cells from rows with data cells. Their proposal works as follows: \begin{inlinenum} \item a row is considered to have meta-data cells if it has at most $50\%$ the average number of cells per row, if it contains no structured cells, or if the visual features are different from the visual features of the others rows; \item then, it tries to detect if the input table is a listing or a matrix; \item once the table structure is identified, it is easy to identify the meta-data. \end{inlinenum} (Note that the authors assume that the input tables are data tables, which is the reason why this cannot be considered a discrimination proposal.)  The authors did not report on their experimental results regarding this task, but their whole system.

\citet{journals/eaai/KimL05} devised a proposal that first attempts to classify the input table. It works as follows: \begin{inlinenum} \item in the case of tables with one single row or column, the first cell is considered to be a meta-data cell and the rest are considered to be data cells; \item in the case of tables with two rows and two columns that do not have any spanned cells, both the first row and column are considered to have meta-data cells and the bottom-right cell is considered to be a data cell; \item tables with two rows/columns and three or more columns/rows whose upper-left cell spans a whole row/column are discarded; otherwise if the first row/colum has some spanned cells (but not all), then the first column/row is assumed to have meta-data cells and the others are assumed to have data cells; \item otherwise, the similarity of the cells is checked per rows and columns using the following functions: a lexical similarity function that focuses on the data types and the length of the contents, and a semantic similarity function that builds on some user-provided key words and patterns. \end{inlinenum}  The authors did not provide any experimental results regarding this task.

\citet{journals/tkde/JungK06} proposed a heuristic-based technique to locate the meta-data within the tables. Their proposal works as follows: \begin{inlinenum} \item cells with a \code{th} tag are assumed to have meta-data; \item if the table can be partitioned into two blocks with the same background colour or font, then the top and/or the left blocks are assumed to contain meta-data; \item if the cells in a row or column have some user-defined contents or match some user-defined patterns, then they are also considered to contain meta-data; \item spanned cells that are embedded in \code{td} tags are also assumed to have meta-data as long as they are located on the top-left areas of the table; its adjacent cells are also considered to have meta-data; \item if the top-right cell is empty, then it is likely that the cells in the first row or column have meta-data; \item a probability is finally computed for every cell building on the previous heuristics and the cells whose probability exceeds a threshold are then considered to be meta-data cells whereas the others are assumed to be data cells. \end{inlinenum}  The evaluation was performed using $10$-fold cross evaluation on $10\,000$ tables from their own repository plus the tables from \cites{conf/das/WangH02} repository.

\citet{journals/pvldb/CafarellaHWWZ08} devised a machine-learning proposal. It works as follows: \begin{inlinenum} \item a learning set is assembled with data tables in which the cells are classified as either meta-data or data cells; \item in cases in which a table does not have any meta-data cells, synthetic cells are created and the meta-data is fed from a separate database with similar tables; \item some structural and content features are computed for each cell; \item a classifier is machine-learnt from the previous features; \item the results of the classifier are used to enrich the other database. \end{inlinenum}  The authors evaluated their proposal by means of $5$-fold cross evaluation on a repository with $1\,000$ tables that were gathered from the Web.

\citet{journals/ijdar/EmbleyKNS16} devised a heuristic-based proposal that searches for four critical cells that help delimit where the meta-data and the data cells are located.  These cells are referred to as CC1, CC2, CC3, and CC4.  CC1 and CC2 identify the top left-most region such that the cells on the right and below that region are mostly meta-data cells; CC3 and CC4 identify the bottom right-most region whose cells are mostly data cells. (Note that CC4 is identified in their segmentation task.)  Their proposal works as follows: \begin{inlinenum} \item it sets CC1 to the top left-most cell and CC2 to the bottom left-most cell; \item it then iteratively shifts CC2 upwards or rightwards while searching for the minimum set of cells between CC1 and CC2 that result in headers that can identify the cells between CC2 and CC4; \item then, CC3 is set to the first cell below to the right of CC2 that does not belong to an empty row or column;  \item after that, footnotes are identified in cells whose contents start with a footnote-mark symbol; \item finally, it analyses some dependencies amongst the meta-data cells to find out the order in which they must be grouped.
\end{inlinenum} The evaluation was performed on a repository with $199$ tables that was provided by \citet{conf/grec/PadmanabhanJKNSS09}.

\citet{conf/nldb/MilosevicGHN16} restricted their attention to tables from the PubMed Central repository. Their focus is on identifying the meta-data cells, since the other cells are considered data cells by default. Their proposal works in three phases.  In the first phase, it searches for \code{thead} tags; if they are found, then the inner \code{th} tags are assumed to encode the meta-data cells and their procedure finishes.  Otherwise, the second phase is intended to find meta-data cells at the top rows as follows: \begin{inlinenum} \item they examine the syntactic similarity of cells on a per column basis; the cells at the top whose syntactic type is different from the cells below, if any, are considered meta-data cells as long as the cells in the same rows in adjacent columns are also considered meta-data cells; \item if a cell in the first row spans several columns, then it is assumed to have meta-data, as well as the cells in the rows below, until a non-spanned cell is found; \item the cells at the top that are between horizontal lines are considered meta-data if they are marked with a \code{thead} tag and they are not empty; in cases in which only one cell in a row has meta-data, the authors refer to it as a super row. \end{inlinenum} The third phase is intended to find meta-data cells on the left columns as follows: \begin{inlinenum} \item the cells on the left-most column that are spanned are meta-data cells and so are the cells on the right until the first non-spanned cell is found; \item the first column below a super row is considered to have meta-data cells that are referred to as stubs. \end{inlinenum} They used a repository with $3\,573$ tables from which $101$ tables were randomly selected to evaluate their proposal.

\subsection{Structural analysis}

\citet{conf/icdar/PennHLM01}, \citet{conf/www/Hurst02}, \citet{conf/das/WangH02}, \citet{journals/eaai/KimL05}, \citet{journals/tkde/JungK06}, \citet{conf/www/GatterbauerBHKP07}, \citet{conf/hci/OkadaM07}, \citet{journals/pvldb/CafarellaHWWZ08}, \citet{conf/wsdm/CrestanP11}, \citet{journals/sigmod/LautertSD13}, \citet{journals/asc/SonP13}, \citet{conf/bdc/EberiusBHTAL15}, \citet{journals/ijdar/EmbleyKNS16}, \citet{conf/aaai/NishidaSHM17}, and \citet{conf/aisc/LiaoLZL18} did not report on any proposals to implement the structural analysis task. \citet{conf/coling/ChenTT00} presented a naive proposal that works on tables that provide data about a single entity, so all of the data cells form a single tuple; regarding the meta-data cells, they group them into headers horizontally or vertically after expanding spanned cells. \citet{conf/wda/YoshidaT01} presented a naive proposal that classifies tables in a number of categories, which makes identifying the tuples quite a trivial task. \citet{journals/vldb/ElmeleegyMH11} also assumed that the tuples within tables that are encoded as lists are always laid out row-wise. \citet{conf/ijcai/LingHWY13} and \citet{conf/er/BraunschweigTL15} assumed that tuples are displayed row-wise or column-wise depending on the number of meta-data or data cells found in the first few rows or columns.  \citet{conf/sigmod/ChuHCG15} also presented a naive approach that assumes that the tuples within tables that are encoded as lists are always laid out row-wise.  \citet{conf/ksem/WuCWFW16} presented an additional naive approach since they just identify tuples in horizontal listings. The other proposals provide more sophisticated approaches.

\citet{conf/ijcai/LermanKM01} used a couple of algorithms to detect row-wise tuples. Their proposal works as follows: \begin{inlinenum} \item first, it uses DataPro to find the patterns that describe the data in each column; \item such patterns can be interpreted as tags that allow to transform a table into a sequence of symbols; \item then, a version of ALERGIA is used to infer a finite automaton from those sequences; \item the automaton is then transformed into a regular expression; \item finally, it identifies repeating sub-patterns that correspond to the tuples in the original table. \end{inlinenum}  No experimentation was performed regarding this task.

\citet{conf/www/CohenHJ02} presented a proposal that relies on four so-called builders, namely: a builder focuses on meta-data cells that cut in on the table, one that focuses on columns of headers, another that focuses on rows of headers, and an additional one that takes the tag paths into account.  The builders are fed into a FOIL-based system in order to learn a classification rule that allows to identify both horizontal and vertical tuples. No experimental results were reported regarding this specific task, but their whole system.

\citet{conf/widm/YangL02} presented a proposal that specialises in numerical tables. It works as follows: \begin{inlinenum} \item first, it removes the headers of the input table; \item then, it checks whether the tuples seem to be one-dimensional or two-dimensional using some heuristics; \item the type of cells is analysed using pre-defined patterns in order to label numeric data cells; \item given the types of cells and the dimensionality of the tuples, their proposal tries to match a number of pre-defined patterns that help identify the tuples. \end{inlinenum} The evaluation was performed on $169$ one-dimensional and $50$ two-dimensional tables.

\citet{conf/sigmod/LermanGMK04} devised two proposals to identify tuples, namely: a constraint solving technique and a probabilistic technique. The former works as follows: \begin{inlinenum} \item it models the cells in the tables using Boolean variables; \item it then adds constraints to ensure that each cell belongs to a single tuple, only contiguous cells can be assigned to the same tuple, and two cells cannot be in the same position in the same table; \item then a constraint solver is used to find a solution to the constraints. \end{inlinenum}  The latter works as follows: \begin{inlinenum} \item it uses a set of observable variables that model the types of tokens in the data cells, and a set of hidden variables, which provide the tuple number or the column number to which every cell belongs; \item a probabilistic model is then learnt by assuming a number of dependencies between token types, cells, columns, neighbour columns, format, or tuple numbers; \item finally, the contents of the hidden variables are inferred building on the probabilistic model. \end{inlinenum}  Their evaluation was performed on the tables from their own repository, which were gathered from $12$ web sites on book sellers, property taxes, white reports, and prisons.

\citet{conf/ieaaie/FumarolaWBMH11} presented a proposal that was described very shallowly.  It seems to work on so-called candidate lists, which are sets of cells that correspond to different columns and form a single tuple; each candidate list is a sub-tree of the DOM tree and they all are required to satisfy some structural similarity constraints, including a minimum size in terms of nodes.  The evaluation was performed on $224$ tables from \cites{conf/www/GatterbauerBHKP07} repository.

\citet{conf/nldb/MilosevicGHN16} identify the tuples according to how the meta-data cells are arranged within a table. If meta-data cells are at the top-most rows and on the left-most columns, then, the table is a matrix with a single tuple that consists of the whole set of data cells; if there are meta-data cells at the top, but not on the left, then the table is a listing in which each row is a tuple; if there are not any meta-data cells, every single data cell corresponds to a tuple. (Note that this proposal cannot be considered a discrimination proposal since it assumes that the input tables are data tables; recall that their focus was on tabled in PubMed publications, which are tables with scientific data.) They used a repository with $3\,573$ tables from which $101$ tables were selected to evaluate their proposal.

\subsection{Interpretation}

\citet{conf/ijcai/LermanKM01}, \citet{conf/icdar/PennHLM01}, \citet{conf/wda/YoshidaT01}, \citet{conf/www/CohenHJ02}, \citet{conf/www/Hurst02}, \citet{conf/das/WangH02}, \citet{conf/sigmod/LermanGMK04}, \citet{journals/eaai/KimL05}, \citet{journals/tkde/JungK06}, \citet{conf/www/GatterbauerBHKP07}, \citet{conf/hci/OkadaM07}, \citet{conf/wsdm/CrestanP11}, \citet{journals/vldb/ElmeleegyMH11}, \citet{conf/ieaaie/FumarolaWBMH11}, \citet{journals/sigmod/LautertSD13}, \citet{conf/ijcai/LingHWY13}, \citet{journals/asc/SonP13}, \citet{conf/er/BraunschweigTL15}, \citet{conf/sigmod/ChuHCG15}, \citet{conf/bdc/EberiusBHTAL15}, \citet{conf/aaai/NishidaSHM17}, and \citet{conf/aisc/LiaoLZL18} did not report on this task.

Most of the other authors reported on naive solutions.  \cites{conf/coling/ChenTT00} proposal works on tables with a single tuple that can spread across several blocks, each of which has its own headers; for each component of the tuple, it creates field descriptors by joining the meta-data cells in the corresponding rows and/or columns.  \cites{journals/ijdar/EmbleyKNS16} proposal is similar, but they focused on tables with a single block.  \citet{conf/nldb/MilosevicGHN16} reported on a naive approach, too: in matrices or listings, they create descriptors for each component from the meta-data in the corresponding column and/or row; in enumerations, they use the caption of the table as a descriptor for every component in the tuples. \citet{conf/widm/YangL02} proposed a similar procedure, but it takes multiple header rows or columns into account, in which case the cells are simply merged to create field descriptors, as well as cells that contain both meta-data and data, in which case the meta-data are transformed into simple descriptors.

The proposal by \citet{journals/pvldb/CafarellaHWWZ08} goes a step forward in cases in which a table does not provide any meta-data cells.  In such cases, they collect the data on a per-column basis and attempt to find the most similar data in the ACSDb database, which is a resource that has many data with correct descriptors. The authors did not report on the evaluation of this task.  \citet{conf/ksem/WuCWFW16} went also a bit further since they used several ad-hoc interpretation methods depending on the structure of the table identified in the discrimination task. They only reported on a method to extract information from horizontal listings with headers using some heuristics that are related to how the \code{th} and the \code{td} tags encode a subject-predicate-object relation. They conducted their experimentation on a repository with $100$ horizontal listings from Wikipedia. The authors evaluated their proposal on a repository with $5\,000$ tables that were randomly selected from the Wikipedia.


\section{Comparison of proposals}
\label{sec:comparison-proposals}

In this section, we compare the proposals that we have summarised in the previous section by means of a comparison framework with both general and task-specific characteristics.

The general characteristics are the following: \begin{inlinenum} \item[Foundation:] it is a hint on the technique behind each proposal. \item[Tables required:] it is the minimum number of tables required for a proposal to work; the less tables required, the better. \item[Effectiveness:] it is the extent to which a proposal succeeds in implementing a task correctly according to an effectiveness measure; the higher the effectiveness, the better. \item[Efficiency:] it is the amount of computing power that a proposal requires to implement a task; the more efficient (i.e., the less computing power is required), the better. \item[Resources:] it refers to the resources that a user must provide so that a proposal can work properly; the less resources, the better. \item[Features:] it refers to the features onto which the input data must be projected in order to machine learn a predictor or to make a decision according to a heuristic.  Features can be either structural, which are related to the HTML or the DOM representation of the input documents, visual, which are related to how they are displayed, or content features, which are related to the contents of the cells. \item[Parameters:] it refers to the settings that must be tuned so that a proposal works well, which can be either pre-defined, learnable, or user-defined parameters.  Pre-defined parameters have a value that the authors of a proposal have found generally appropriate; they are preferable to learnable parameters, whose values must be experimentally learnt by the user; in turn, they are preferable to user-defined parameters, which must be set by the user using his or her intuition; the less parameters, the better. \end{inlinenum}

Note that it is easy to make decisions building on the general features that we presented above since we have characterised their preferred contents; the same applies to the task-specific features that we describe in the following sub-sections. The only exceptions are the foundation characteristic and the features characteristic.  The reason is that it is not generally clear whether a heuristic-based approach is preferable to a machine-learning approach or vice versa, or whether structural, visual, or content features are preferable to each other. Note, too, that effectiveness and efficiency are decision-making characteristics, but the figures provided by an author are not generally comparable to the figures provided by a different author because they evaluated their proposals using different approaches, learning sets, evaluation sets, and machinery.

\subsection{Location}

\begin{table*}
    \begin{adjustbox}{angle=90}
    	\includegraphics[width=63em]{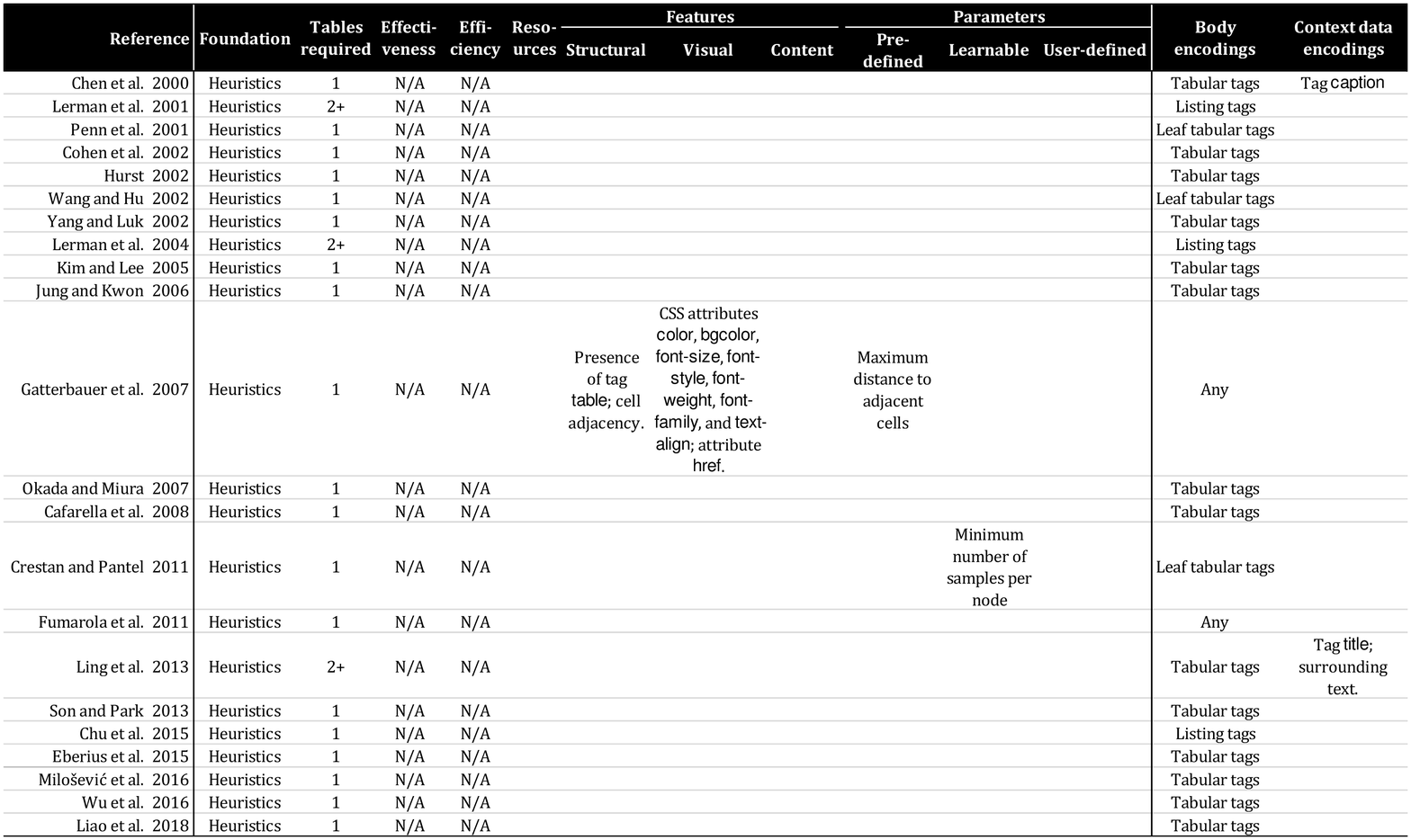}
    \end{adjustbox}
	\caption{Comparison of location proposals.}
	\label{tab:comparison-location-proposals}
\end{table*}

Table~\ref{tab:comparison-location-proposals} summarises our comparison regarding location proposals.  The task-specific characteristics are the following: \begin{inlinenum} \item[Body encodings:] it refers to how the tables that a proposal can locate must be encoded; the more kinds of encodings are identified, the better. \item[Context-data encodings:] it refers to how context-data cells are encoded; the more kinds of encodings are identified, the better. \end{inlinenum}

Regarding the general characteristics, it is surprising that all of the location proposals are based on heuristics; there is no record in the literature of a single proposal that has tried a machine-learning approach. Most proposals can work on a single table, but the ones by \citet{conf/ijcai/LermanKM01, conf/sigmod/LermanGMK04} and \citet{conf/ijcai/LingHWY13} require at least a pair of tables to perform table alignment. None of the proposals was presented in isolation, but as a component of a larger system, which is the reason why no author reported on effectiveness or efficiency.  Realise that only the proposal by \citet{conf/www/GatterbauerBHKP07} projects the input documents onto structural and visual features in order to apply their heuristics; note, too, that it is the only one that requires a pre-defined parameter.  The proposal by \citet{conf/wsdm/CrestanP11} is the only that requires the user to set a learnable parameter.

Regarding the task-specific characteristics, most of the proposals locate tables that are encoded using tabular tags, a few focus on tables that are encoded using listing tags, and only \cites{conf/www/GatterbauerBHKP07} and \cites{conf/ieaaie/FumarolaWBMH11} proposals are independent from the tags used since they analyse how the input documents are displayed. Note, too, that the vast majority of proposals focus on locating the tables themselves, not their context data. \citet{conf/coling/ChenTT00} and \citet{conf/ijcai/LingHWY13} are the exceptions: the former presents a simple approach that searches for \code{caption} tags and the latter presents a more sophisticated approach that analyses the \code{title} tags and the text that surrounds the tables.

\subsection{Segmentation}

\begin{table*}
    \begin{adjustbox}{angle=90}
    	\includegraphics[width=63em]{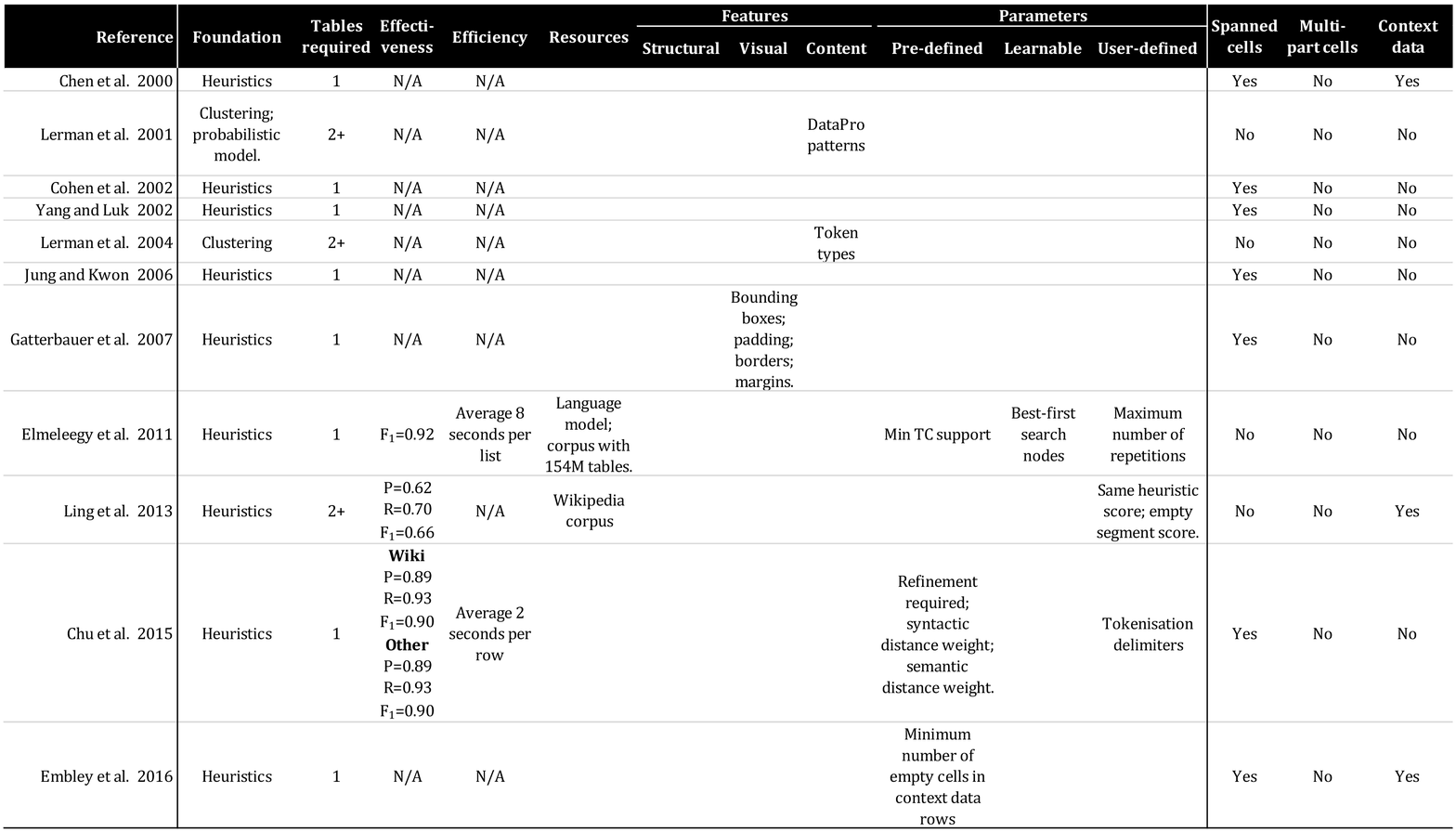}
    \end{adjustbox}
    \caption{Comparison of segmentation proposals.}
	\label{tab:comparison-segmentation-proposals}
\end{table*}

Table~\ref{tab:comparison-segmentation-proposals} summarises our comparison regarding segmentation proposals.  The task-specific characteristics are the following: \begin{inlinenum} \item[Spanned cells:] it describes if a proposal is able to identify cells that span multiple columns and/or rows; a proposal that can identify spanned cells is better than a proposal that cannot. \item[Multi-part cells:] it describes if a proposal is able to identify cells that provide partial contents and must be merged; a proposal that can identify multi-part cells is better than a proposal that cannot. \item[Context data:] it describes if a proposal can identify context data or not; a proposal that can identify context data is better than a proposal that cannot. \end{inlinenum}

Regarding the general characteristics, it is easy to realise that only the proposals by \citet{conf/ijcai/LermanKM01, conf/sigmod/LermanGMK04} have tried machine-learning approaches; the others rely on heuristics that their authors have proven to work well in practice. Furthermore, most of them can work on as few as one input table, but the ones by \citet{conf/ijcai/LermanKM01, conf/sigmod/LermanGMK04} and \citet{conf/ijcai/LingHWY13}.  Unfortunately, roughly $72\%$ of the authors did not report on the effectiveness of their proposals; the others reported on precision, recall, and/or the $F_1$ score. Only \citet{journals/vldb/ElmeleegyMH11} and \citet{conf/sigmod/ChuHCG15} reported on the efficiency of their approaches; their figures reveal that the algorithms behind the scenes might not be scalable enough. Regarding the resources required, only the proposals by \citet{journals/vldb/ElmeleegyMH11} and \citet{conf/ijcai/LingHWY13} require the user to provide a few, but they do not seem to be difficult to find. Only the proposals by \citet{conf/ijcai/LermanKM01, conf/sigmod/LermanGMK04} and \citet{conf/www/GatterbauerBHKP07} require to project the input tables onto some simple features.  Regarding the parameters, only the proposals by \citet{journals/vldb/ElmeleegyMH11}, \citet{conf/ijcai/LingHWY13}, and \citet{conf/sigmod/ChuHCG15} need the users to set a few.

Regarding the task-specific characteristics, it is surprising that many proposals do not make an attempt to analyse spanned cells and that none of them can identify multi-part cells, both of which are very common in practice. It is also surprising that only the proposals by \citet{conf/coling/ChenTT00}, \citet{conf/ijcai/LingHWY13}, and \citet{journals/ijdar/EmbleyHLN06} can identify context data, which are also very common in practice; unfortunately, the proposal by \citet{conf/coling/ChenTT00} cannot be considered a general solution to the problem since it is very naive.

\subsection{Discrimination}

\begin{table*}
    \begin{adjustbox}{angle=90}
    	\includegraphics[width=63em]{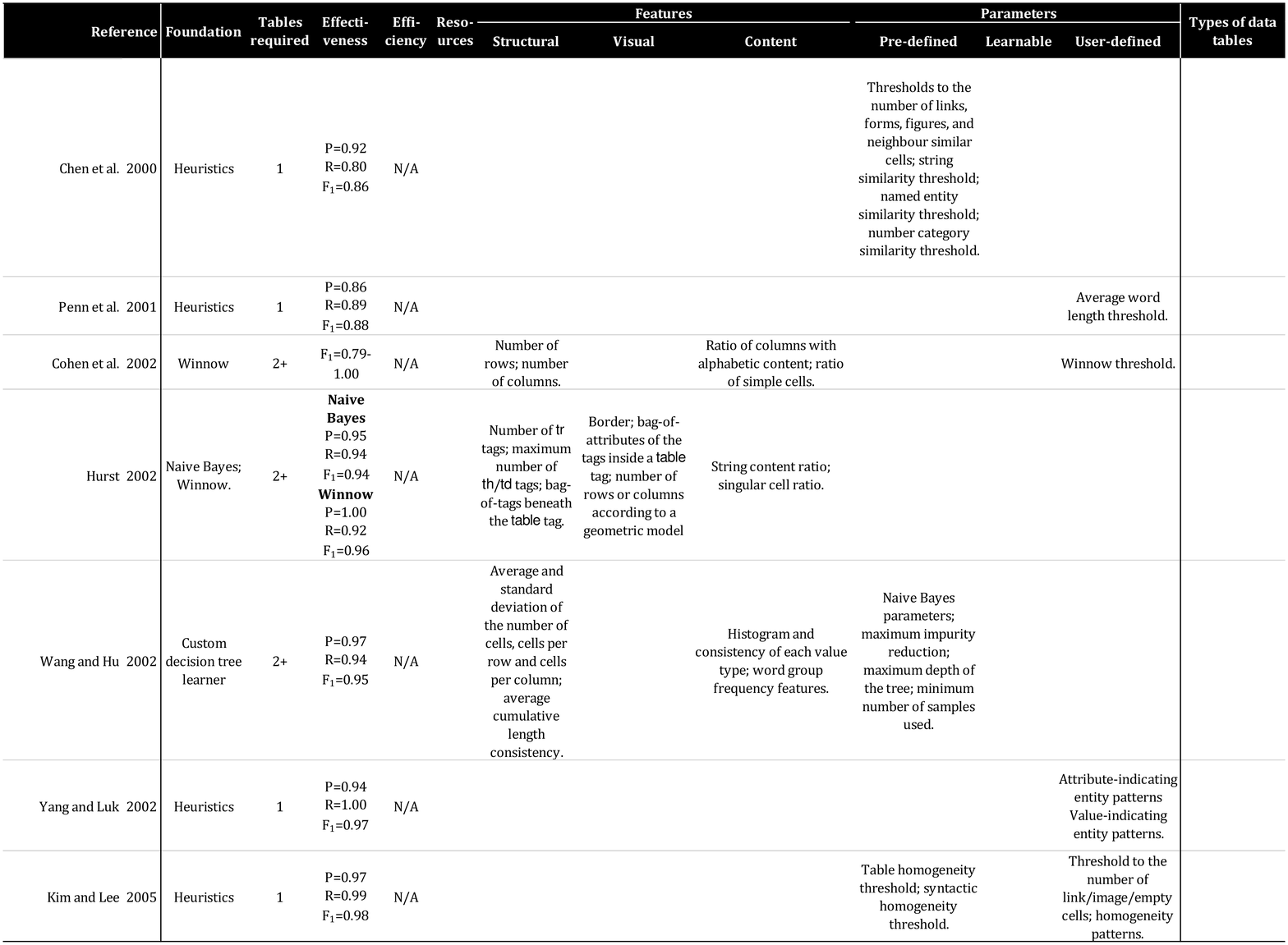}
    \end{adjustbox}	
	\caption{Comparison of discrimination proposals (Part 1).}
	\label{tab:comparison-discrimination-proposals-1}
\end{table*}

\begin{table*}
    \begin{adjustbox}{angle=90}
    	\includegraphics[width=63em]{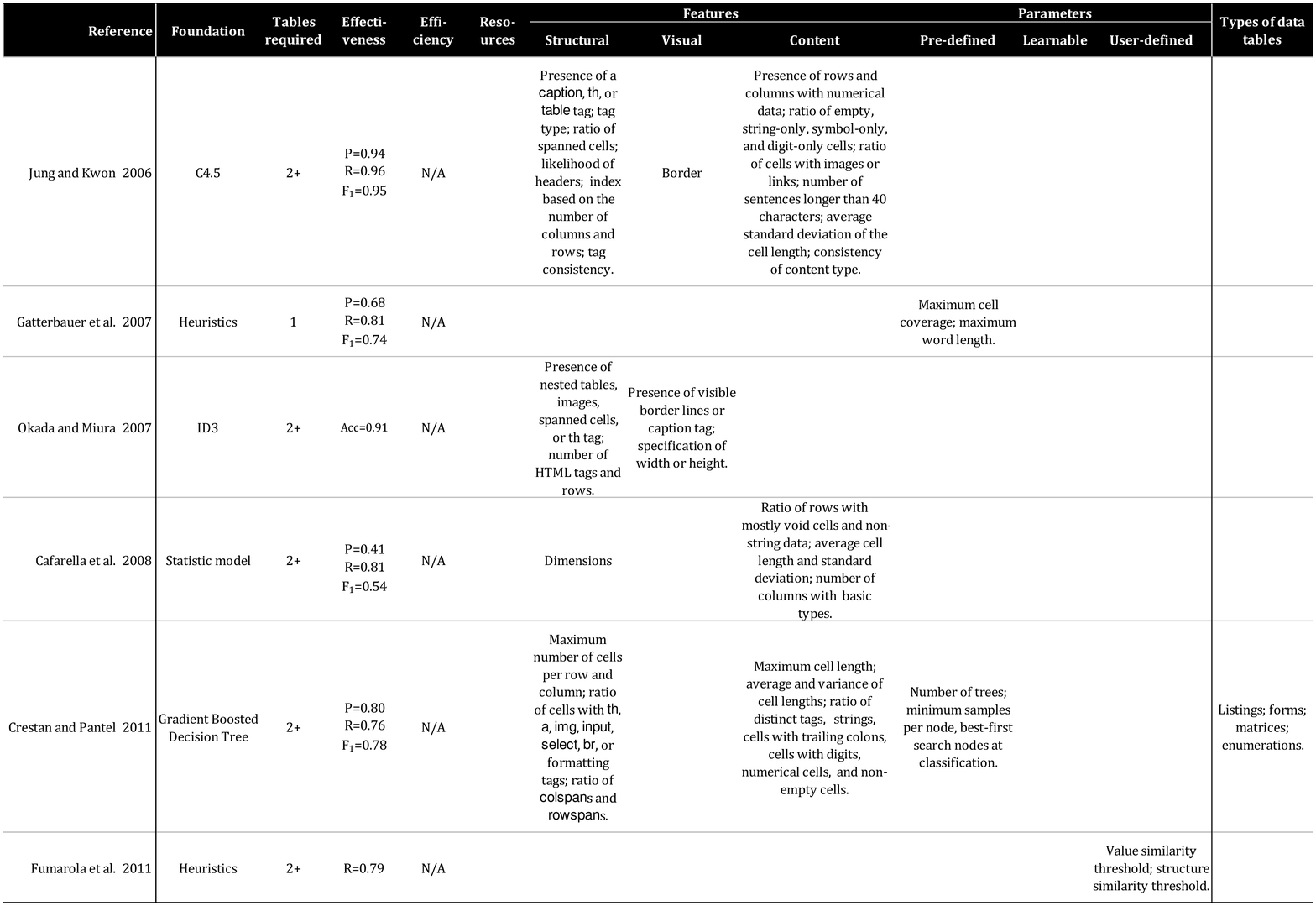}
    \end{adjustbox}	
    \caption{Comparison of discrimination proposals (Part 2).}
	\label{tab:comparison-discrimination-proposals-2}
\end{table*}

\begin{table*}
    \begin{adjustbox}{angle=90}
    	\includegraphics[width=63em]{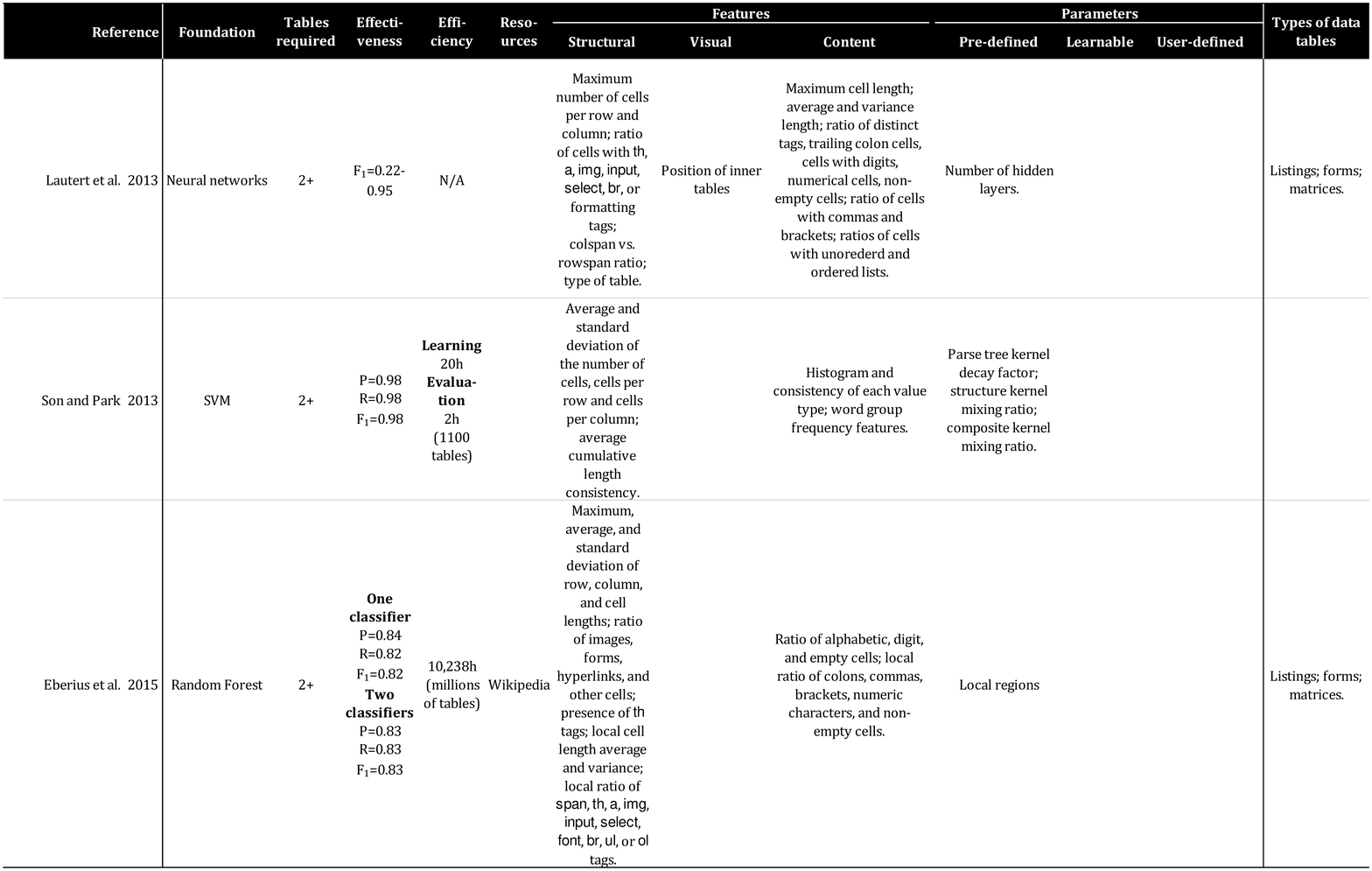}
    \end{adjustbox}	
    \caption{Comparison of discrimination proposals (Part 3).}
	\label{tab:comparison-discrimination-proposals-3}
\end{table*}

\begin{table*}
    \begin{adjustbox}{angle=90}
    	\includegraphics[width=63em]{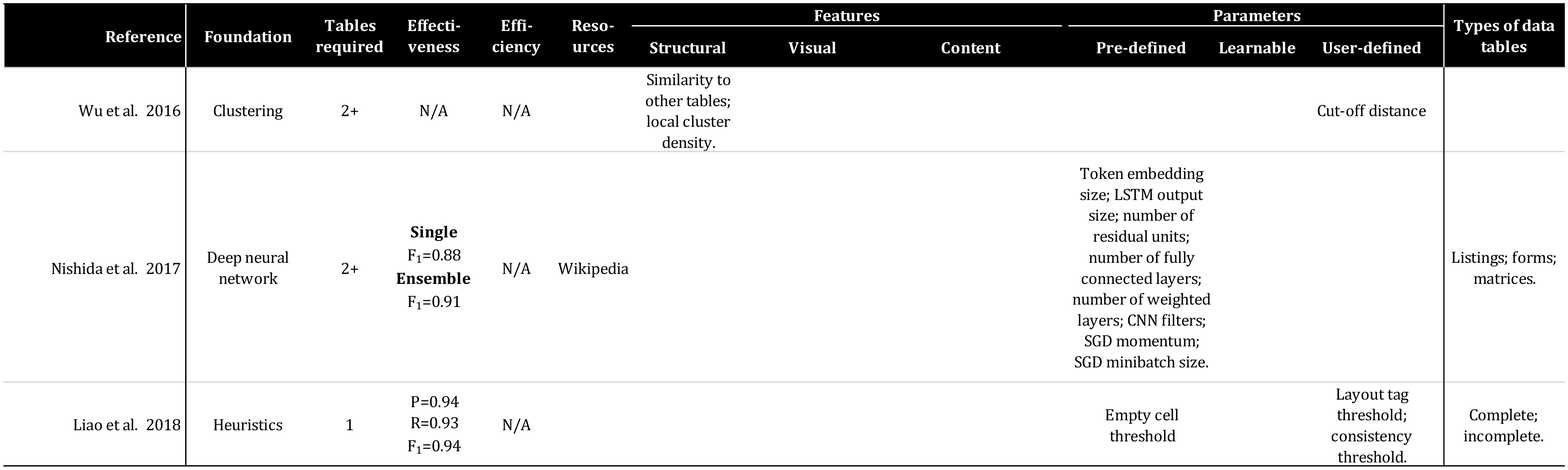}
    \end{adjustbox}	
    \caption{Comparison of discrimination proposals (Part 4).}
	\label{tab:comparison-discrimination-proposals-4}
\end{table*}

Tables~\ref{tab:comparison-discrimination-proposals-1}--\ref{tab:comparison-discrimination-proposals-4} summarise our comparison regarding discrimination proposals.  The only task-specific characteristic is \xitem[Types of data tables], which refers to the kinds of data tables that a proposal can discriminate; the more types can be discriminated, the better.

Regarding the general characteristics, it is easy to realise that $64\%$ of the proposals use a machine-learning approach and the rest use heuristic-based approaches.  The former require at least two tables to learn a predictor that implements the discrimination task, whereas the latter can generally work on a single table.  Except for \cites{conf/wecwis/WuCY02}, the other authors report on effectiveness measures that are specific to this task; most of the authors selected precision, recall, and the $F_1$ score as effectiveness measures; the exceptions are \citet{conf/www/CohenHJ02}, \citet{journals/sigmod/LautertSD13}, and \citet{conf/aaai/NishidaSHM17}, who report on the $F_1$ score only, \citet{conf/hci/OkadaM07}, who reported on accuracy, and \citet{conf/ieaaie/FumarolaWBMH11}, who reported on recall only. Apparently, the effectiveness of the machine-learning proposals is higher than the effectiveness of the heuristic-based proposals; however, due to the differences in the evaluation processes, this conclusion is not sound.  Unfortunately, only \citet{journals/asc/SonP13} and \citet{conf/bdc/EberiusBHTAL15} reported on the efficiency of their proposals, which does not seem to be very good according to their figures; \citet{conf/wecwis/WuCY02} did not report on the efficiency of their proposal but they mentioned that it relies on a linear clustering algorithm. The only proposals that require resources are the ones by \citet{conf/bdc/EberiusBHTAL15} and \citet{conf/aaai/NishidaSHM17}; fortunately, they do not seem to be a major obstacle since they consists in a corpus that was gathered from the Wikipedia.  The ones that rely on machine learning project the input data onto a space of structural, visual, and/or content features that seem simple to compute. Regarding their parameters, most of them have pre-defined parameters for which the authors recommend some values that are expected work generally well; none of the proposals require any learnable parameters, but a few require user-defined parameters.

Regarding the task-specific characteristics, the only proposals that can sub-classify data tables are the following ones: \citet{conf/wsdm/CrestanP11} distinguishes amongst listings, forms, matrices, and enumerations; \citet{journals/sigmod/LautertSD13}, \citet{conf/bdc/EberiusBHTAL15}, and \citet{conf/aaai/NishidaSHM17} distinguish amongst listings, forms, and matrices; and \citet{conf/aisc/LiaoLZL18} distinguishes between complete and incomplete tables (which are encoded as independent tables, but must be stitched together so that they can be properly interpreted).

\subsection{Functional analysis}

\begin{table*}
    \vskip2em
    \begin{adjustbox}{angle=90}
        \includegraphics[width=61em]{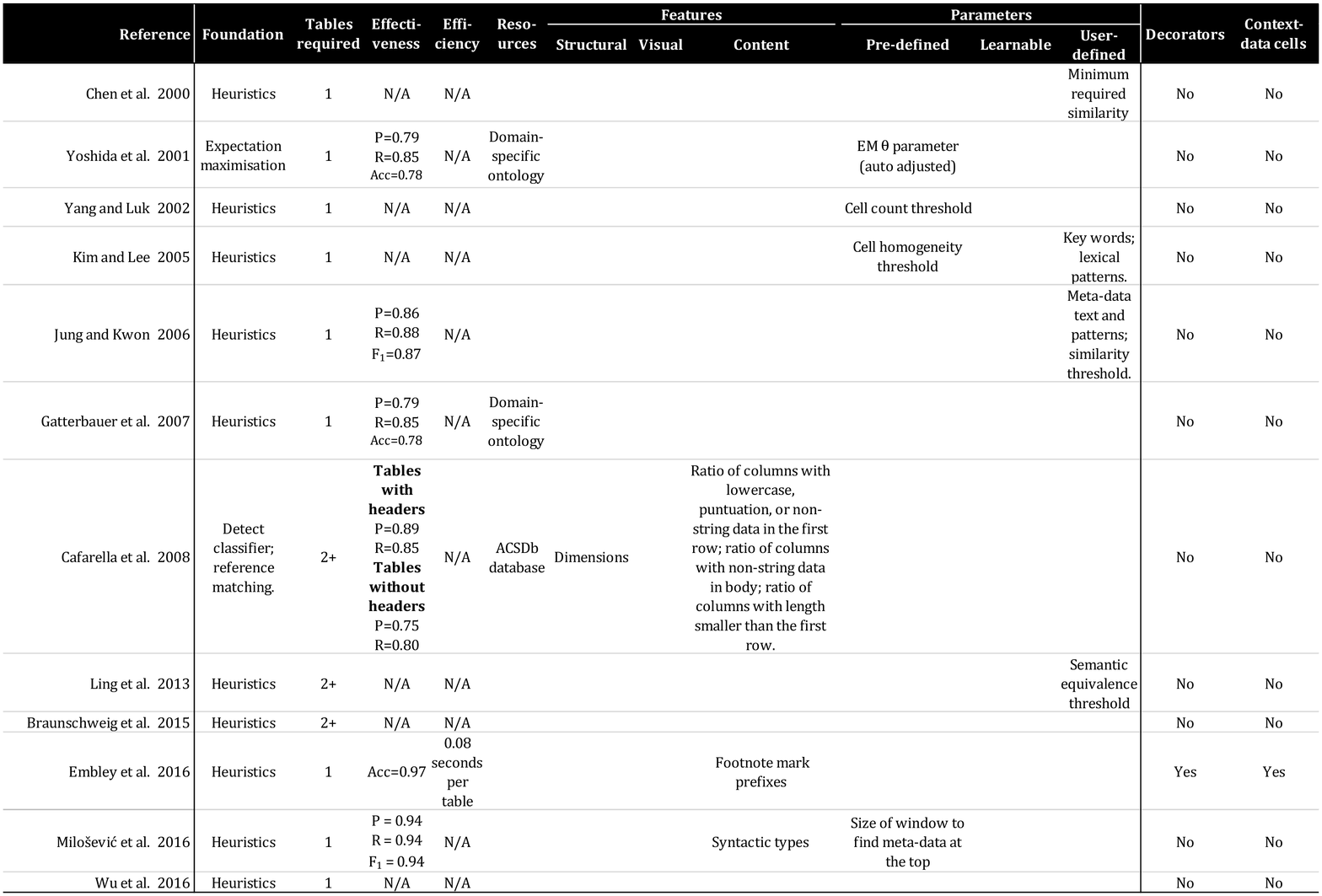}
    \end{adjustbox}		
	\caption{Comparison of functional analysis proposals.}
	\label{tab:comparison-functional-analysis-proposals}
\end{table*}

Table~\ref{tab:comparison-functional-analysis-proposals} summarises our comparison regarding functional analysis proposals.

The task-specific characteristics are the following: \begin{inlinenum} \item[Context-data cells:] it describes if a proposal is able to make context-data cells apart from the others; a proposal that can identify context-data cells is better than a proposal that cannot. \xitem[Decorators:] it refers to the ability of a proposal to identify decorator cells; a proposal that can find decorators is better than a proposal that cannot. \end{inlinenum}

Regarding the general characteristics, $83\%$ of the proposals rely on heuristics and rest rely on machine-learning approaches.  Most of them can work on as few as a single table, with the exception of the proposals by \citet{journals/pvldb/CafarellaHWWZ08}, \citet{conf/coling/ChenTT00}, and \citet{conf/coling/ChenTT00}, we need to compare at least two tables.  Many of the authors report on the effectiveness of their proposals; realise that most of the measures are below $0.90$, which means that there is enough room for improvement regarding this task.  Unfortunately, only \citet{journals/ijdar/EmbleyHLN06} reported on the efficiency of their proposal, which seems scalable enough. Regarding the resources required, \cites{conf/wda/YoshidaT01} and \cites{conf/www/GatterbauerBHKP07} proposals require domain-specific ontologies, whereas \cites{conf/webdb/CafarellaHZWW08} requires a publicly-available database. The proposal by \citet{conf/webdb/CafarellaHZWW08} is the only that projects the input data onto a space of simple structural and content features.  The proposal by \citet{conf/wda/YoshidaT01} requires a pre-defined parameter that is auto-adjusted, and the proposals by \citet{conf/widm/YangL02}, \citet{journals/eaai/KimL05}, and \citet{conf/nldb/MilosevicGHN16} require another pre-defined parameter for which the authors provide a default value; the only proposals that require user-defined parameters are the ones by \citet{conf/coling/ChenTT00}, \citet{journals/eaai/KimL05}, \citet{journals/tkde/JungK06}, and \citet{conf/ijcai/LingHWY13}.

Regarding the task-specific characteristics, note that only the proposal by \citet{conf/nldb/MilosevicGHN16} can identify some decorator cells and context-data cells.  This is a bit surprising since, according to our experience, these kinds of cells are very common in practice.

\subsection{Structural analysis}

\begin{table*}
    \begin{adjustbox}{angle=90}
        \includegraphics[width=63em]{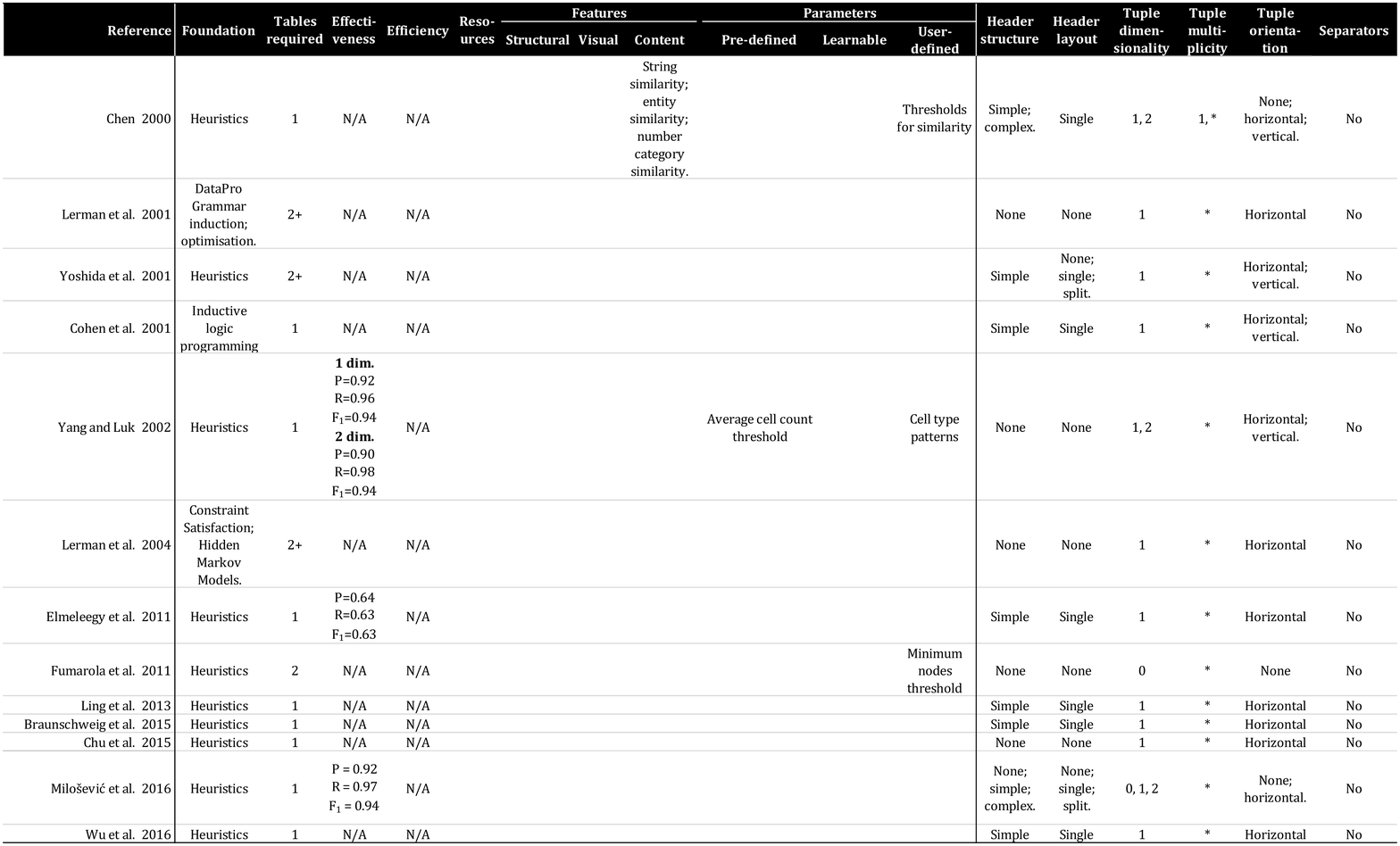}
    \end{adjustbox}		
    \caption{Comparison of structural analysis proposals.}
	\label{tab:comparison-structural-analysis-proposals}
\end{table*}

Table~\ref{tab:comparison-structural-analysis-proposals} summarises our comparison regarding structural analysis proposals.  The task-specific characteristics are the following: \begin{inlinenum} \item[Header structure:] it describes the kinds of headers that a proposal can identify according to their structure, namely: \code{none}, which means that it can analyse tables without headers, \code{simple}, which means that it can analyse simple headers that consists of one meta-data cell only, and \code{complex}, which means that it can identify complex headers that consists of multiple meta-data cells; the more header structures a proposal can identify, the better.  \item[Header layout:] it describes the kinds of headers that a proposal can identify according to how they are laid out, namely: \code{none}, which means that it can identify that a table does not have any headers, \code{single}, which means that it can identify headers in the first rows and/or columns of a table, \code{horizontally repeated}, which means that it can identify that the headers are repeated every some rows, \code{vertically repeated}, which means that it can identify that the headers are repeated every some columns, and \code{split}, which means that it can identify series of headers that are split across several non-adjacent rows or columns; the more header layouts a proposal can identify, the better. \item[Tuple dimensionality:] it describes the dimensionality of the tuples that a proposal can identify, namely: \code{0} if it can identify the tuples in an enumeration, \code{1} it can identify the tuples in a listing or a form, and \code{2} if it can identify the tuple in a matrix; the more tuple dimensionalities a proposal can identify, the better. \item[Tuple multiplicity:] it describes the number of tuples that a table is intended to show, namely: \code{1} in the case of forms and matrices, and \code{*} in the case of listings and enumerations; the more tuple multiplicities a proposal can identify, the better. \item[Tuple orientation:] it describes the orientations that it can identify, namely: \code{none} in the case of matrices and enumerations, \code{horizontal} or \code{vertical} in the case of listings and forms; the more tuple orientations a proposal can identify, the better.  \item[Separators:] it describes whether a proposal can identify separator rows and/or columns; a proposal that can identify separators is better than a proposal that cannot. \end{inlinenum}

Regarding the general characteristics, many proposals rely on heuristic-based approaches; the exceptions are the proposals by \citet{conf/ijcai/LermanKM01, conf/sigmod/LermanGMK04}, which leverage some grammar induction techniques, and \cites{conf/www/CohenHJ02} proposal, which leverages inductive logic programming. Most of the proposals require as few as one input table; the exceptions are the proposals by \citet{conf/ijcai/LermanKM01, conf/sigmod/LermanGMK04}, \citet{conf/wda/YoshidaT01}, and \citet{conf/ieaaie/FumarolaWBMH11}, which require two tables for comparison purposes.  Unfortunately, only \citet{conf/widm/YangL02}, \citet{journals/vldb/ElmeleegyMH11}, and \citet{conf/nldb/MilosevicGHN16} reported on the effectiveness of their proposals, and none of the authors reported on their efficiency. Note that none of the proposals require to project the input data onto a space of features, but the one by \citet{conf/coling/ChenTT00}.  Note, too, that \citet{conf/coling/ChenTT00}, \cites{conf/widm/YangL02}, and \cites{conf/ieaaie/FumarolaWBMH11} proposals are the only that have parameters.

Regarding the task-specific characteristics, it is surprising that most of the proposals assume that the tables do not have any headers or they are simple, except for \cites{conf/nldb/MilosevicGHN16} proposal; it is also surprising that the only proposal that can identify single and split headers is the one by \citet{conf/wda/YoshidaT01}. Regarding the tuple dimensionality, only the proposals by \citet{conf/widm/YangL02} and \citet{conf/nldb/MilosevicGHN16} can make uni-dimensional tuples apart from two-dimensional tuples; \cites{conf/nldb/MilosevicGHN16} can also deal with zero-dimensional tuples; the proposal by \citet{conf/ieaaie/FumarolaWBMH11} implicitly assumes that the tuples in a table are zero-dimensional and does not make an attempt to analyse the structure of the corresponding cells; the other proposals implicitly assume that the tuples are uni-dimensional.  Regarding the tuple multiplicity, it is interesting to see that all of the proposals assume that tables may display more than one tuple; simply put, they cannot make listings apart from forms.  Regarding the tuple orientation, most proposals implicitly assume that the tuples are oriented horizontally; the only exceptions are the proposals by \citet{conf/wda/YoshidaT01}, \citet{conf/www/CohenHJ02}, and \citet{conf/widm/YangL02}, which can make horizontal tuples apart from vertical tuples. It is surprising that none of the proposals that we have surveyed can identify separators, even though they are very common in practice.

\subsection{Interpretation}

\begin{table*}
    \begin{adjustbox}{angle=90}
        \includegraphics[width=63em]{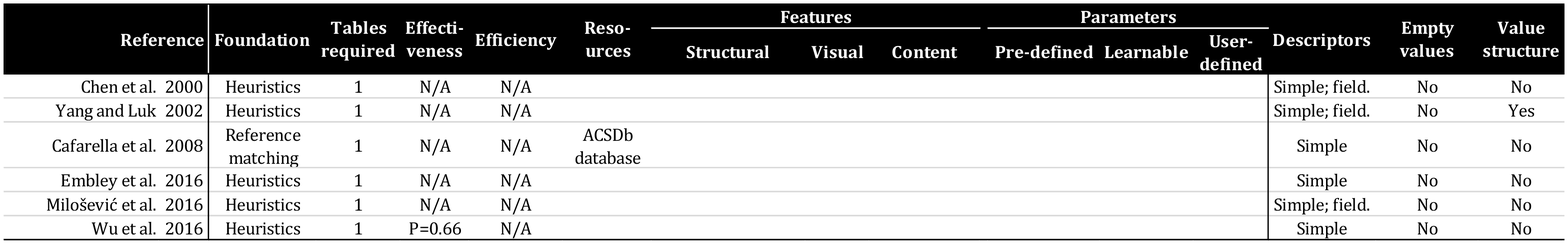}
    \end{adjustbox}		
	\caption{Comparison of interpretation proposals.}
	\label{tab:comparison-interpretation-proposals}
\end{table*}

Table~\ref{tab:comparison-interpretation-proposals} summarises our comparison regarding interpretation proposals.  The task-specific characteristics are the following: \begin{inlinenum} \item[Descriptors:] it reports on the kind of descriptors that a proposal can assign to the data in a table; the more kinds of descriptors a proposal can generate, the better. \item[Empty contents:]  it refers to the ability of a proposal to make a difference between empty cells whose contents are factorised and cells that are actually empty; a proposal that can make a difference between factorised and void cells is better than another proposal that cannot. \item[Content structure:] it refers to the ability of a proposal to make a difference between atomic cells and structured cells; a proposal that can make a difference between atomic cells and structured cells is better than a proposal that cannot. \end{inlinenum}

Regarding the general characteristics, most proposals rely on heuristics that have proven to work well in practice; the only exception is the proposal by \citet{conf/webdb/CafarellaHZWW08}, which uses a reference matching approach. \citet{conf/wecwis/WuCY02} were the only authors who reported on effectiveness, but they measured precision only; unfortunately, none of the proposals report on efficiency.  \cites{conf/webdb/CafarellaHZWW08} proposal is the only one that requires a publicly-available resource. None of the proposals project the input data onto a feature space and none of them require any parameters to be set.

Regarding the task-specific characteristics, all of the proposals can generate simple descriptors; only the proposals by \citet{conf/coling/ChenTT00}, \citet{conf/widm/YangL02}, and \citet{conf/nldb/MilosevicGHN16} can generate field descriptors.  Unfortunately, none of the proposals can make a difference between factorised cells and void cells. Regarding making a difference amongst atomic and structured cells, it seems that only the proposal by \citet{conf/widm/YangL02} can deal with this problem.


\section{Conclusions}
\label{sec:conclusions}

This article summarises and compares many proposals that have been published between $2000$ and $2018$ regarding extracting data from tables that are encoded using HTML.  The problem is not trivial insofar many tables are encoded using a subset of table-related tags that help locate and segment them, but do not provide a clue on the function of the cells or their structure; many others are encoded using listing tags, block tags, or other tags that look like a table when they are displayed, which hampers locating and segmenting them.

Our analysis makes it clear that none of the proposals that we have listed provide a complete solution to the data-extraction problem.  Most of them address only some of the tasks involved and they differ regarding the problems that they address within each task.  Regarding the location task, most proposals focus on tables that are encoded using table-related tags, there are a couple that focus on listing tags, and also a couple that are independent from the tags used; what seems an actual challenge is to identify context data, since the few proposals that take this problem into account are very naive. Regarding the segmentation task, it is surprising that no proposal can identify multi-part cells and that most of them do not attempt to segment the context data. The discrimination task is the one that has been paid more attention, but not many proposals attempt to go further than making non-data tables apart from data tables; recent proposals attempt to classify data tables in more categories since this definitely helps interpret them. Regarding the functional analysis task, it is surprising that almost none of the proposals pay attention to identifying context-data cells or decorators cells. Regarding the structural analysis task, the problems that have got none or very little attention are identifying split headers and zero- and two-dimensional tuples.  Regarding the interpretation tasks, creating artificial descriptors in cases in which not enough meta-data are available, analysing whether an empty value is actually empty or factorised, and analysing the structure of the contents of a cell are problems that have not been addressed sufficiently.  Addressing these problems would help expand the kinds of tables from which data can be extracted.

Last, but clearly not least, the evaluation of the proposals is also a very relevant problem.  We have found that many authors have used \cites{conf/das/WangH02} repository in addition to their own repositories; unfortunately, the subsets of tables selected were different and their sizes range from as many as $342\,795$ tables to a hundred tables or less.  Definitely, recent repositories like DWDTC~\cite{conf/bdc/EberiusBHTAL15} or WDC~\cite{conf/www/LehmbergRMB16} will help.  We have also found many authors who used $k$-fold cross evaluation, but there is not a general consensus; there is not even a consensus regarding the value of $k$ in the cases in which this procedure was used.  As a conclusion, the experimental results reported are not comparable to each other.  Neither is it common to find figures regarding efficiency, which makes it difficult to realise if a proposal might work well in a production scenario.  \citet{journals/kbs/JimenezCS16} set a foundation regarding how to evaluate information extraction proposals in general, but they did not focus on the tasks involved in extracting information from tables that are encoded using HTML.

Summing up: extracting data from tables that are encoded in HTML is an active research field in which we expect new results to be published in the near future. We hope that this article helps researchers sift through the state-of-the-art proposals in this field.


\section*{Acknowledgments}

The work by Juan C.\@ Roldán, Patricia Jiménez, and Rafael Corchuelo was supported by the Spanish R\&D programme with grants TIN2013-40848-R and TIN2016-75394-R.  The work by Juan C.\@ Roldán was also supported by the Fulbright programme.

    \bibliographystyle{abbrvnat}
    \bibliography{Bibliography}

\begin{thebibliography}{72}
\providecommand{\natexlab}[1]{#1}
\providecommand{\url}[1]{\texttt{#1}}
\expandafter\ifx\csname urlstyle\endcsname\relax
  \providecommand{\doi}[1]{doi: #1}\else
  \providecommand{\doi}{doi: \begingroup \urlstyle{rm}\Url}\fi

\bibitem[Braunschweig et~al.(2015)Braunschweig, Thiele, and
  Lehner]{conf/er/BraunschweigTL15}
K.~Braunschweig, M.~Thiele, and W.~Lehner.
\newblock From web tables to concepts: a semantic normalization approach.
\newblock In \emph{ER}, pages 247--260, 2015.
\newblock \doi{10.1007/978-3-319-25264-3_18}.

\bibitem[Buchsbaum et~al.(2000)Buchsbaum, Caldwell, Church, Fowler, and
  Muthukrishnan]{conf/soda/BuchsbaumCCFM00}
A.~L. Buchsbaum, D.~F. Caldwell, K.~W. Church, G.~S. Fowler, and
  S.~Muthukrishnan.
\newblock Engineering the compression of massive tables: an experimental
  approach.
\newblock In \emph{SODA}, pages 175--184, 2000.
\newblock URL \url{http://dl.acm.org/citation.cfm?id=338219.338249}.

\bibitem[Cafarella et~al.(2008{\natexlab{a}})Cafarella, Halevy, Wang, Wu, and
  Zhang]{journals/pvldb/CafarellaHWWZ08}
M.~J. Cafarella, A.~Y. Halevy, D.~Z. Wang, E.~Wu, and Y.~Zhang.
\newblock {W}eb{T}ables: exploring the power of tables on the {W}eb.
\newblock \emph{PVLDB}, 1\penalty0 (1):\penalty0 538--549, 2008{\natexlab{a}}.
\newblock URL \url{http://www.vldb.org/pvldb/1/1453916.pdf}.

\bibitem[Cafarella et~al.(2008{\natexlab{b}})Cafarella, Halevy, Zhang, Wang,
  and Wu]{conf/webdb/CafarellaHZWW08}
M.~J. Cafarella, A.~Y. Halevy, Y.~Zhang, D.~Z. Wang, and E.~Wu.
\newblock Uncovering the relational {W}eb.
\newblock In \emph{WebDB}, 2008{\natexlab{b}}.
\newblock URL
  \url{http://webdb2008.como.polimi.it/images/stories/WebDB2008/paper30.pdf}.

\bibitem[Cafarella et~al.(2018)Cafarella, Halevy, Lee, Madhavan, Yu, Wang, and
  Wu]{journals/pvldb/CafarellaHLMYWW18}
M.~J. Cafarella, A.~Y. Halevy, H.~Lee, J.~Madhavan, C.~Yu, D.~Z. Wang, and
  E.~Wu.
\newblock Ten years of web tables.
\newblock \emph{PVLDB}, 11\penalty0 (12):\penalty0 2140--2149, 2018.
\newblock \doi{10.14778/3229863.3240492}.

\bibitem[Cannaviccio et~al.(2018)Cannaviccio, Ariemma, Barbosa, and
  Merialdo]{conf/webdb/CannaviccioABM18}
M.~Cannaviccio, L.~Ariemma, D.~Barbosa, and P.~Merialdo.
\newblock Leveraging wikipedia table schemas for knowledge graph augmentation.
\newblock In \emph{WebDB}, pages 5:1--5:6, 2018.
\newblock \doi{10.1145/3201463.3201468}.

\bibitem[Chen et~al.(2000)Chen, Tsai, and Tsai]{conf/coling/ChenTT00}
H.-H. Chen, S.-C. Tsai, and J.-H. Tsai.
\newblock Mining tables from large scale {HTML} texts.
\newblock In \emph{COLING}, pages 166--172, 2000.
\newblock URL \url{http://aclweb.org/anthology/C00-1025}.

\bibitem[Christen(2012)]{books/springer/Christen12}
P.~Christen.
\newblock \emph{Data Matching - Concepts and Techniques for Record Linkage,
  Entity Resolution, and Duplicate Detection}.
\newblock Springer, 2012.
\newblock \doi{10.1007/978-3-642-31164-2}.

\bibitem[Chu et~al.(2015{\natexlab{a}})Chu, He, Chakrabarti, and
  Ganjam]{conf/sigmod/ChuHCG15}
X.~Chu, Y.~He, K.~Chakrabarti, and K.~Ganjam.
\newblock {TEGRA}: table extraction by global record alignment.
\newblock In \emph{SIGMOD}, pages 1713--1728, 2015{\natexlab{a}}.
\newblock \doi{10.1145/2723372.2723725}.

\bibitem[Chu et~al.(2015{\natexlab{b}})Chu, Morcos, Ilyas, Ouzzani, Papotti,
  Tang, and Ye]{conf/sigmod/ChuMIOP0Y15}
X.~Chu, J.~Morcos, I.~F. Ilyas, M.~Ouzzani, P.~Papotti, N.~Tang, and Y.~Ye.
\newblock {KATARA}: a data cleaning system powered by knowledge bases and
  crowdsourcing.
\newblock In \emph{SIGMOD Conference}, pages 1247--1261, 2015{\natexlab{b}}.
\newblock \doi{10.1145/2723372.2749431}.

\bibitem[Cimmino and Corchuelo(2018{\natexlab{a}})]{conf/bis/CimminoC18}
A.~Cimmino and R.~Corchuelo.
\newblock On feeding business systems with linked resources from the {W}eb of
  {D}ata.
\newblock In \emph{BIS}, pages 307--320, 2018{\natexlab{a}}.
\newblock \doi{10.1007/978-3-319-93931-5_22}.

\bibitem[Cimmino and Corchuelo(2018{\natexlab{b}})]{conf/hais/CimminoC18}
A.~Cimmino and R.~Corchuelo.
\newblock A hybrid genetic-bootstrapping approach to link resources in the
  {W}eb of {D}ata.
\newblock In \emph{HAIS}, pages 145--157, 2018{\natexlab{b}}.
\newblock \doi{10.1007/978-3-319-92639-1_13}.

\bibitem[Cohen et~al.(2002)Cohen, Hurst, and Jensen]{conf/www/CohenHJ02}
W.~W. Cohen, M.~Hurst, and L.~S. Jensen.
\newblock A flexible learning system for wrapping tables and lists in {HTML}
  documents.
\newblock In \emph{WWW}, pages 232--241, 2002.
\newblock \doi{10.1145/511446.511477}.

\bibitem[Costa-Silva et~al.(2006)Costa-Silva, Jorge, and
  Torgo]{journals/ijdar/SilvaJT06}
A.~Costa-Silva, A.~M. Jorge, and L.~Torgo.
\newblock Design of an end-to-end method to extract information from tables.
\newblock \emph{IJDAR}, 8\penalty0 (2-3):\penalty0 144--171, 2006.
\newblock \doi{10.1007/s10032-005-0001-x}.

\bibitem[Crestan and Pantel(2010)]{conf/cikm/CrestanP10}
E.~Crestan and P.~Pantel.
\newblock A fine-grained taxonomy of tables on the {W}eb.
\newblock In \emph{CIKM}, pages 1405--1408, 2010.
\newblock \doi{10.1145/1871437.1871633}.

\bibitem[Crestan and Pantel(2011)]{conf/wsdm/CrestanP11}
E.~Crestan and P.~Pantel.
\newblock Web-scale table census and classification.
\newblock In \emph{WSDM}, pages 545--554, 2011.
\newblock \doi{10.1145/1935826.1935904}.

\bibitem[Dong et~al.(2014)Dong, Gabrilovich, Heitz, Horn, Lao, Murphy,
  Strohmann, Sun, and Zhang]{conf/kdd/DongGHHLMSSZ14}
X.~Dong, E.~Gabrilovich, G.~Heitz, W.~Horn, N.~Lao, K.~Murphy, T.~Strohmann,
  S.~Sun, and W.~Zhang.
\newblock Knowledge vault: a web-scale approach to probabilistic knowledge
  fusion.
\newblock In \emph{KDD}, pages 601--610, 2014.
\newblock \doi{10.1145/2623330.2623623}.

\bibitem[Eberius et~al.(2015)Eberius, Braunschweig, Hentsch, Thiele, Ahmadov,
  and Lehner]{conf/bdc/EberiusBHTAL15}
J.~Eberius, K.~Braunschweig, M.~Hentsch, M.~Thiele, A.~Ahmadov, and W.~Lehner.
\newblock Building the {D}resden {W}eb {T}able {C}orpus: a classification
  approach.
\newblock In \emph{BDC}, pages 41--50, 2015.
\newblock \doi{10.1109/BDC.2015.30}.

\bibitem[Efthymiou et~al.(2017)Efthymiou, Hassanzadeh, Rodr{\'\i}guez-Muro, and
  Christophides]{conf/semweb/EfthymiouHRC17}
V.~Efthymiou, O.~Hassanzadeh, M.~Rodr{\'\i}guez-Muro, and V.~Christophides.
\newblock Matching web tables with knowledge base entities: from entity lookups
  to entity embeddings.
\newblock In \emph{ISWC}, pages 260--277, 2017.
\newblock \doi{10.1007/978-3-319-68288-4_16}.

\bibitem[Elmeleegy et~al.(2011)Elmeleegy, Madhavan, and
  Halevy]{journals/vldb/ElmeleegyMH11}
H.~Elmeleegy, J.~Madhavan, and A.~Y. Halevy.
\newblock Harvesting relational tables from lists on the {W}eb.
\newblock \emph{VLDB}, 20\penalty0 (2):\penalty0 209--226, 2011.
\newblock \doi{10.1007/s00778-011-0223-0}.

\bibitem[Embley et~al.(2006)Embley, Hurst, Lopresti, and
  Nagy]{journals/ijdar/EmbleyHLN06}
D.~W. Embley, M.~Hurst, D.~P. Lopresti, and G.~Nagy.
\newblock Table-processing paradigms: a research survey.
\newblock \emph{IJDAR}, 8\penalty0 (2-3):\penalty0 66--86, 2006.
\newblock \doi{10.1007/s10032-006-0017-x}.

\bibitem[Embley et~al.(2016)Embley, Krishnamoorthy, Nagy, and
  Seth]{journals/ijdar/EmbleyKNS16}
D.~W. Embley, M.~S. Krishnamoorthy, G.~Nagy, and S.~C. Seth.
\newblock Converting heterogeneous statistical tables on the {W}eb to
  searchable databases.
\newblock \emph{IJDAR}, 19\penalty0 (2):\penalty0 119--138, 2016.
\newblock \doi{10.1007/s10032-016-0259-1}.

\bibitem[Fan et~al.(2014)Fan, Lu, Ooi, Tan, and Zhang]{conf/icde/FanLOTZ14}
J.~Fan, M.~Lu, B.~C. Ooi, W.-C. Tan, and M.~Zhang.
\newblock A hybrid machine-crowdsourcing system for matching web tables.
\newblock In \emph{ICDE}, pages 976--987, 2014.
\newblock \doi{10.1109/ICDE.2014.6816716}.

\bibitem[Fumarola et~al.(2011)Fumarola, Weninger, Barber, Malerba, and
  Han]{conf/ieaaie/FumarolaWBMH11}
F.~Fumarola, T.~Weninger, R.~Barber, D.~Malerba, and J.~Han.
\newblock Extracting general lists from web documents: a hybrid approach.
\newblock In \emph{IEAAIE}, pages 285--294, 2011.
\newblock \doi{10.1007/978-3-642-21822-4_29}.

\bibitem[Galkin et~al.(2015)Galkin, Mouromtsev, and Auer]{conf/kesw/GalkinMA15}
M.~Galkin, D.~Mouromtsev, and S.~Auer.
\newblock Identifying web tables: supporting a neglected type of content on the
  {W}eb.
\newblock In \emph{KESW}, pages 48--62, 2015.
\newblock \doi{10.1007/978-3-319-24543-0_4}.

\bibitem[Gatterbauer et~al.(2007)Gatterbauer, Bohunsky, Herzog, Kr{\"u}pl, and
  Pollak]{conf/www/GatterbauerBHKP07}
W.~Gatterbauer, P.~Bohunsky, M.~Herzog, B.~Kr{\"u}pl, and B.~Pollak.
\newblock Towards domain-independent information extraction from web tables.
\newblock In \emph{WWW}, pages 71--80, 2007.
\newblock \doi{10.1145/1242572.1242583}.

\bibitem[Hurst(2001)]{conf/wda/Hurst01}
M.~Hurst.
\newblock Layout and language: challenges for table understanding on the {W}eb.
\newblock In \emph{WDA}, pages 27--30, 2001.
\newblock URL \url{http://wda2001.csc.liv.ac.uk/Papers/12_hurst_wda2001.pdf}.

\bibitem[Hurst(2002)]{conf/www/Hurst02}
M.~Hurst.
\newblock Classifying {TABLE} elements in {HTML}.
\newblock In \emph{WWW}, 2002.
\newblock URL
  \url{http://wwwconference.org/proceedings/www2002/poster/115/index.html}.

\bibitem[Jim{\'e}nez et~al.(2016)Jim{\'e}nez, Corchuelo, and
  Sleiman]{journals/kbs/JimenezCS16}
P.~Jim{\'e}nez, R.~Corchuelo, and H.~A. Sleiman.
\newblock {ARIEX}: automated ranking of information extractors.
\newblock \emph{Knowl.-Based Syst.}, 93:\penalty0 84--108, 2016.
\newblock \doi{10.1016/j.knosys.2015.11.004}.

\bibitem[Jung and Kwon(2006)]{journals/tkde/JungK06}
S.-W. Jung and H.-C. Kwon.
\newblock A scalable hybrid approach for extracting head components from web
  tables.
\newblock \emph{IEEE Trans. Knowl. Data Eng.}, 18\penalty0 (2):\penalty0
  174--187, 2006.
\newblock \doi{10.1109/TKDE.2006.19}.

\bibitem[Khayyat et~al.(2015)Khayyat, Ilyas, Jindal, Madden, Ouzzani, Papotti,
  Quian{\'{e}}{-}Ruiz, Tang, and Yin]{conf/sigmod/KhayyatIJMOPQ0Y15}
Z.~Khayyat, I.~F. Ilyas, A.~Jindal, S.~Madden, M.~Ouzzani, P.~Papotti,
  J.~Quian{\'{e}}{-}Ruiz, N.~Tang, and S.~Yin.
\newblock {B}ig{D}ansing: a system for big data cleansing.
\newblock In \emph{SIGMOD Conference}, pages 1215--1230, 2015.
\newblock \doi{10.1145/2723372.2747646}.

\bibitem[Kim and Lee(2005)]{journals/eaai/KimL05}
Y.-S. Kim and K.-H. Lee.
\newblock Detecting tables in web documents.
\newblock \emph{Eng. Appl. of AI}, 18\penalty0 (6):\penalty0 745--757, 2005.
\newblock \doi{10.1016/j.engappai.2005.01.009}.

\bibitem[Knoblock et~al.(2017)Knoblock, Szekely, Fink, Degler, Newbury,
  Sanderson, Blanch, Snyder, Chheda, Jain, Krishna, Sreekanth, and
  Yao]{conf/semweb/KnoblockSFDNSBS17}
C.~A. Knoblock, P.~A. Szekely, E.~E. Fink, D.~Degler, D.~Newbury, R.~Sanderson,
  K.~Blanch, S.~Snyder, N.~Chheda, N.~Jain, R.~R. Krishna, N.~B. Sreekanth, and
  Y.~Yao.
\newblock Lessons learned in building linked data for the {A}merican {A}rt
  {C}ollaborative.
\newblock In \emph{ISWC}, pages 263--279, 2017.
\newblock \doi{10.1007/978-3-319-68204-4_26}.

\bibitem[Lautert et~al.(2013)Lautert, Scheidt, and
  Dorneles]{journals/sigmod/LautertSD13}
L.~R. Lautert, M.~M. Scheidt, and C.~F. Dorneles.
\newblock Web table taxonomy and formalization.
\newblock \emph{SIGMOD Record}, 42\penalty0 (3):\penalty0 28--33, 2013.
\newblock \doi{10.1145/2536669.2536674}.

\bibitem[Lehmberg et~al.(2016)Lehmberg, Ritze, Meusel, and
  Bizer]{conf/www/LehmbergRMB16}
O.~Lehmberg, D.~Ritze, R.~Meusel, and C.~Bizer.
\newblock A large public corpus of web tables containing time and context
  meta-data.
\newblock In \emph{WWW}, pages 75--76, 2016.
\newblock \doi{10.1145/2872518.2889386}.

\bibitem[Lerman et~al.(2001)Lerman, Knoblock, and
  Minton]{conf/ijcai/LermanKM01}
K.~Lerman, C.~Knoblock, and S.~Minton.
\newblock Automatic data extraction from lists and tables in web sources.
\newblock In \emph{IJCAI}, 2001.
\newblock URL \url{http://www.isi.edu/integration/papers/lerman01-atem.pdf}.

\bibitem[Lerman et~al.(2004)Lerman, Getoor, Minton, and
  Knoblock]{conf/sigmod/LermanGMK04}
K.~Lerman, L.~Getoor, S.~Minton, and C.~A. Knoblock.
\newblock Using the structure of web sites for automatic segmentation of
  tables.
\newblock In \emph{SIGMOD}, pages 119--130, 2004.
\newblock \doi{10.1145/1007568.1007584}.

\bibitem[Liao et~al.(2018)Liao, Liu, Zhang, and Liu]{conf/aisc/LiaoLZL18}
T.~Liao, T.~Liu, S.~Zhang, and Z.~Liu.
\newblock Research on web table positioning technology based on table structure
  and heuristic rules.
\newblock In \emph{AISC}, pages 351--360, 2018.
\newblock \doi{10.1007/978-3-319-67071-3_41}.

\bibitem[Ling et~al.(2013)Ling, Halevy, Wu, and Yu]{conf/ijcai/LingHWY13}
X.~Ling, A.~Y. Halevy, F.~Wu, and C.~Yu.
\newblock Synthesizing union tables from the {W}eb.
\newblock In \emph{IJCAI}, 2013.
\newblock URL
  \url{http://www.aaai.org/ocs/index.php/IJCAI/IJCAI13/paper/view/6758}.

\bibitem[Lo et~al.(2000)Lo, Wu, and Yu]{conf/icdcs/LoWY00}
M.-L. Lo, K.-L. Wu, and P.~S. Yu.
\newblock {TabSum}: a flexible and dynamic table summarization approach.
\newblock In \emph{ICDCS}, pages 628--635, 2000.
\newblock \doi{10.1109/ICDCS.2000.840979}.

\bibitem[Lopresti and Nagy(1999)]{conf/iaprgr/LoprestiN99}
D.~P. Lopresti and G.~Nagy.
\newblock Automated table processing: an (opinionated) survey.
\newblock In \emph{GREC}, pages 109--134, 1999.
\newblock URL
  \url{http://www.cse.lehigh.edu/~lopresti/Publications/1999/grec99.pdf}.

\bibitem[Lopresti and Nagy(2000)]{conf/grec/LoprestiN99}
D.~P. Lopresti and G.~Nagy.
\newblock A tabular survey of automated table processing.
\newblock In \emph{GREC}, pages 93--120, 2000.
\newblock \doi{10.1007/3-540-40953-X_9}.

\bibitem[Mankoff et~al.(2005)Mankoff, Fait, and Tran]{conf/chi/MankoffFT05}
J.~Mankoff, H.~Fait, and T.~Tran.
\newblock Is your web page accessible? {A} comparative study of methods for
  assessing web page accessibility for the blind.
\newblock In \emph{CHI}, pages 41--50, 2005.
\newblock \doi{10.1145/1054972.1054979}.

\bibitem[Milo\v{s}evi\'{c} et~al.(2016)Milo\v{s}evi\'{c}, Gregson, Hernandez,
  and Nenadic]{conf/nldb/MilosevicGHN16}
N.~Milo\v{s}evi\'{c}, C.~Gregson, R.~Hernandez, and G.~Nenadic.
\newblock Disentangling the structure of tables in scientific literature.
\newblock In \emph{NLDB}, pages 162--174, 2016.
\newblock \doi{10.1007/978-3-319-41754-7_14}.

\bibitem[Mulwad et~al.(2010)Mulwad, Finin, Syed, and
  Joshi]{conf/semweb/MulwadFSJ10a}
V.~Mulwad, T.~Finin, Z.~Syed, and A.~Joshi.
\newblock Using {L}inked {D}ata to interpret tables.
\newblock In \emph{COLD}, 2010.
\newblock URL \url{http://ceur-ws.org/Vol-665/MulwadEtAl_COLD2010.pdf}.

\bibitem[Nishida et~al.(2017)Nishida, Sadamitsu, Higashinaka, and
  Matsuo]{conf/aaai/NishidaSHM17}
K.~Nishida, K.~Sadamitsu, R.~Higashinaka, and Y.~Matsuo.
\newblock Understanding the semantic structures of tables with a hybrid deep
  neural network architecture.
\newblock In \emph{AAAI}, pages 168--174, 2017.
\newblock URL \url{http://aaai.org/ocs/index.php/AAAI/AAAI17/paper/view/14396}.

\bibitem[Okada and Miura(2007)]{conf/hci/OkadaM07}
H.~Okada and T.~Miura.
\newblock Detection of layout-purpose table tags based on machine learning.
\newblock In \emph{UAHCI}, pages 116--123, 2007.
\newblock \doi{10.1007/978-3-540-73283-9_14}.

\bibitem[Padmanabhan et~al.(2009)Padmanabhan, Jandhyala, Krishnamoorthy, Nagy,
  Seth, and Silversmith]{conf/grec/PadmanabhanJKNSS09}
R.~K. Padmanabhan, R.~C. Jandhyala, M.~S. Krishnamoorthy, G.~Nagy, S.~C. Seth,
  and W.~Silversmith.
\newblock Interactive conversion of web tables.
\newblock In \emph{GREC}, pages 25--36, 2009.
\newblock \doi{10.1007/978-3-642-13728-0_3}.

\bibitem[Penn et~al.(2001)Penn, Hu, Luo, and McDonald]{conf/icdar/PennHLM01}
G.~Penn, J.~Hu, H.~Luo, and R.~T. McDonald.
\newblock Flexible web document analysis for delivery to narrow-bandwidth
  devices.
\newblock In \emph{ICDAR}, pages 1074--1078, 2001.
\newblock \doi{10.1109/ICDAR.2001.953951}.

\bibitem[Peterson(2014)]{books/or/Peterson14}
C.~Peterson.
\newblock \emph{Learning Responsive Web Design}.
\newblock O'Reilly, 2014.

\bibitem[Pimplikar and Sarawagi(2012)]{journals/pvldb/PimplikarS12}
R.~Pimplikar and S.~Sarawagi.
\newblock Answering table queries on the {W}eb using column keywords.
\newblock \emph{PVLDB}, 5\penalty0 (10):\penalty0 908--919, 2012.
\newblock \doi{10.14778/2336664.2336665}.

\bibitem[Qi et~al.(2017)Qi, Wu, and Wang]{conf/IEEEwisa/QiWW17}
F.~Qi, X.~Wu, and N.~Wang.
\newblock Building top-$k$ consistent results for web table augmentation.
\newblock In \emph{WISA}, pages 74--79, 2017.
\newblock \doi{10.1109/WISA.2017.30}.

\bibitem[Ratinov et~al.(2011)Ratinov, Roth, Downey, and
  Anderson]{conf/acl/RatinovRDA11}
L.-A. Ratinov, D.~Roth, D.~Downey, and M.~Anderson.
\newblock Local and global algorithms for disambiguation to {W}ikipedia.
\newblock In \emph{ACL}, pages 1375--1384, 2011.
\newblock URL \url{http://www.aclweb.org/anthology/P11-1138}.

\bibitem[Ren et~al.(2017)Ren, Wu, He, Qu, Voss, Ji, Abdelzaher, and
  Han]{conf/www/RenWHQVJAH17}
X.~Ren, Z.~Wu, W.~He, M.~Qu, C.~R. Voss, H.~Ji, T.~F. Abdelzaher, and J.~Han.
\newblock {C}o{T}ype: joint extraction of typed entities and relations with
  knowledge bases.
\newblock In \emph{WWW}, pages 1015--1024, 2017.
\newblock \doi{10.1145/3038912.3052708}.

\bibitem[Ritze and Bizer(2017)]{conf/edbt/RitzeB17}
D.~Ritze and C.~Bizer.
\newblock Matching web tables to {DB}pedia: a feature utility study.
\newblock In \emph{EDBT}, pages 210--221, 2017.
\newblock \doi{10.5441/002/edbt.2017.20}.

\bibitem[Sarma et~al.(2012)Sarma, Fang, Gupta, Halevy, Lee, Wu, Xin, and
  Yu]{conf/sigmod/SarmaFGHLWXY12}
A.~D. Sarma, L.~Fang, N.~Gupta, A.~Y. Halevy, H.~Lee, F.~Wu, R.~Xin, and C.~Yu.
\newblock Finding related tables.
\newblock In \emph{SIGMOD}, pages 817--828, 2012.
\newblock \doi{10.1145/2213836.2213962}.

\bibitem[Sekhavat et~al.(2014)Sekhavat, Paolo, Barbosa, and
  Merialdo]{conf/www/SekhavatPBM14}
Y.~A. Sekhavat, F.~D. Paolo, D.~Barbosa, and P.~Merialdo.
\newblock Knowledge base augmentation using tabular data.
\newblock In \emph{LDOW}, 2014.
\newblock URL \url{http://ceur-ws.org/Vol-1184/ldow2014_paper_02.pdf}.

\bibitem[Sierra et~al.(2008)Sierra, Fern{\'a}ndez-Valmayor, and
  Fern{\'a}ndez-Manj{\'o}n]{journals/software/SierraFF08}
J.~L. Sierra, A.~Fern{\'a}ndez-Valmayor, and B.~Fern{\'a}ndez-Manj{\'o}n.
\newblock From documents to applications using markup languages.
\newblock \emph{IEEE Software}, 25\penalty0 (2):\penalty0 68--76, 2008.
\newblock \doi{10.1109/MS.2008.36}.

\bibitem[Son and Park(2013)]{journals/asc/SonP13}
J.~W. Son and S.-B. Park.
\newblock Web table discrimination with composition of rich structural and
  content information.
\newblock \emph{Appl. Soft Comput.}, 13\penalty0 (1):\penalty0 47--57, 2013.
\newblock \doi{10.1016/j.asoc.2012.07.025}.

\bibitem[Taheriyan et~al.(2016)Taheriyan, Knoblock, Szekely, and
  Ambite]{journals/ws/TaheriyanKSA16}
M.~Taheriyan, C.~A. Knoblock, P.~A. Szekely, and J.~L. Ambite.
\newblock Learning the semantics of structured data sources.
\newblock \emph{J. Web Semant.}, 37-38:\penalty0 152--169, 2016.
\newblock \doi{10.1016/j.websem.2015.12.003}.

\bibitem[Taleb et~al.(2015)Taleb, Dssouli, and Serhani]{conf/bigdata/TalebDS15}
I.~Taleb, R.~Dssouli, and M.~A. Serhani.
\newblock {B}ig {D}ata pre-processing: a quality framework.
\newblock In \emph{IEEE Intl. Congress on Big Data}, pages 191--198, 2015.
\newblock \doi{10.1109/BigDataCongress.2015.35}.

\bibitem[Tschirschnitz et~al.(2017)Tschirschnitz, Papenbrock, and
  Naumann]{journals/tods/TschirschnitzPN17}
F.~Tschirschnitz, T.~Papenbrock, and F.~Naumann.
\newblock Detecting inclusion dependencies on very many tables.
\newblock \emph{ACM Trans. Database Syst.}, 42\penalty0 (3):\penalty0
  18:1--18:29, 2017.
\newblock \doi{10.1145/3105959}.

\bibitem[Venetis et~al.(2011)Venetis, Halevy, Madhavan, Pasca, Shen, Wu, Miao,
  and Wu]{journals/pvldb/VenetisHMPSWMW11}
P.~Venetis, A.~Y. Halevy, J.~Madhavan, M.~Pasca, W.~Shen, F.~Wu, G.~Miao, and
  C.~Wu.
\newblock Recovering semantics of tables on the {W}eb.
\newblock \emph{PVLDB}, 4\penalty0 (9):\penalty0 528--538, 2011.
\newblock \doi{10.14778/2002938.2002939}.

\bibitem[Wang and Hu(2002)]{conf/das/WangH02}
Y.~Wang and J.~Hu.
\newblock Detecting tables in {HTML} documents.
\newblock In \emph{DAS}, pages 249--260, 2002.
\newblock \doi{10.1007/3-540-45869-7_29}.

\bibitem[Wu et~al.(2002)Wu, Chen, and Yu]{conf/wecwis/WuCY02}
K.-L. Wu, S.-K. Chen, and P.~S. Yu.
\newblock Dynamic refinement of table summarization for m-commerce.
\newblock In \emph{WECWIS}, pages 179--186, 2002.
\newblock \doi{10.1109/WECWIS.2002.1021257}.

\bibitem[Wu et~al.(2016)Wu, Cao, Wang, Fu, and Wang]{conf/ksem/WuCWFW16}
X.~Wu, C.~Cao, Y.~Wang, J.~Fu, and S.~Wang.
\newblock Extracting knowledge from web tables based on {DOM} tree similarity.
\newblock In \emph{KSEM}, pages 302--313, 2016.
\newblock \doi{10.1007/978-3-319-47650-6_24}.

\bibitem[Yakout et~al.(2012)Yakout, Ganjam, Chakrabarti, and
  Chaudhuri]{conf/sigmod/YakoutGCC12}
M.~Yakout, K.~Ganjam, K.~Chakrabarti, and S.~Chaudhuri.
\newblock Infogather: entity augmentation and attribute discovery by holistic
  matching with web tables.
\newblock In \emph{SIGMOD Conference}, pages 97--108, 2012.
\newblock \doi{10.1145/2213836.2213848}.

\bibitem[Yang and Luk(2002)]{conf/widm/YangL02}
Y.~Yang and W.-S. Luk.
\newblock A framework for web table mining.
\newblock In \emph{WIDM}, pages 36--42, 2002.
\newblock \doi{10.1145/584931.584940}.

\bibitem[Yoshida et~al.(2001)Yoshida, Torisawa, and
  Tsujii]{conf/wda/YoshidaT01}
M.~Yoshida, K.~Torisawa, and J.~Tsujii.
\newblock A method to integrate tables of the {W}orld {W}ide {W}eb.
\newblock In \emph{WDA}, pages 31--34, 2001.
\newblock URL \url{http://wda2001.csc.liv.ac.uk/Papers/13_yoshida_wda2001.pdf}.

\bibitem[Zanibbi et~al.(2004)Zanibbi, Blostein, and
  Cordy]{journals/ijdar/ZanibbiBC04}
R.~Zanibbi, D.~Blostein, and J.~R. Cordy.
\newblock A survey of table recognition.
\newblock \emph{IJDAR}, 7\penalty0 (1):\penalty0 1--16, 2004.
\newblock \doi{10.1007/s10032-004-0120-9}.

\bibitem[Zhang and Chakrabarti(2013)]{conf/sigmod/ZhangC13}
M.~Zhang and K.~Chakrabarti.
\newblock {I}nfo{G}ather+: semantic matching and annotation of numeric and
  time-varying attributes in web tables.
\newblock In \emph{SIGMOD Conference}, pages 145--156, 2013.
\newblock \doi{10.1145/2463676.2465276}.

\bibitem[Zhang et~al.(2013)Zhang, Chen, Chen, Du, and
  Zou]{conf/dasfaa/ZhangCCDZ13}
X.~Zhang, Y.~Chen, J.~Chen, X.~Du, and L.~Zou.
\newblock Mapping entity-attribute web tables to web-scale knowledge bases.
\newblock In \emph{DASFAA}, pages 108--122, 2013.
\newblock \doi{10.1007/978-3-642-37450-0_8}.

\end{thebibliography}

\end{document}